\documentclass[12pt,preprint]{emulateapj}
\usepackage{lscape}
\begin{document}

\title{The Role of Mass and Environment in Multiple Star Formation: A 2MASS
Survey of Wide Multiplicity in Three Young Associations}

\author{Adam L. Kraus (alk@astro.caltech.edu), Lynne A. Hillenbrand (lah@astro.caltech.edu)}
\affil{California Institute of Technology, Department of Astrophysics, MC 105-24, Pasadena, CA 91125}

\begin{abstract}

We present the results of a search for wide binary systems among 783 members 
of three nearby young associations: Taurus-Auriga, Chamaeleon I, and two 
subgroups of Upper Scorpius.  Near-infrared ($JHK$) imagery from 2MASS was 
analyzed to search for wide (1-30\arcsec;  $\sim$150-4500 AU) companions to 
known association members, using color-magnitude cuts to reject likely 
background stars. We identify a total of 131 candidate binary companions with 
colors consistent with physical association, of which 39 have not been 
identified previously in the literature. Our results suggest that the wide 
binary frequency is a function of both mass and environment, with 
significantly higher frequencies among high-mass stars than lower-mass stars 
and in the T associations than in the OB association. We discuss the 
implications for wide binary formation and conclude that the environmental 
dependence is not a direct result of stellar density or total association 
mass, but instead might depend on another environmental parameter like the gas 
temperature.  We also analyze the mass ratio distribution as a function of 
mass and find that it agrees with the distribution for field stars to within 
the statistical uncertainties. The binary populations in these associations 
generally follow the empirical mass-maximum separation relation observed for 
field binaries, but we have found one candidate low-mass system 
(USco-160611.9-193532; $M_{tot}$$\sim$0.4 $M_{\sun}$) which has a projected 
separation (10.8\arcsec; 1550 AU) much larger than the suggested limit for its 
mass. Finally, we find that the binary frequency in the USco-B subgroup is 
significantly higher than in the USco-A subgroup and is consistent with the 
measured values in Taurus and Cham I. This discrepancy, the absence of 
high-mass stars in USco-B, and its marginally distinct kinematics suggest that 
it might not be directly associated with the OB associations of Sco-Cen, but 
instead represents an older analogue of the younger $\rho$ Oph or Lupus 
associations.

\end{abstract}


\section{Introduction}

The frequency and properties of multiple star systems are important diagnostics 
for placing constraints on star formation processes and calibrating stellar 
evolutionary models. This has prompted numerous attempts to characterize the 
properties of nearby binary systems in the field. Multiplicity surveys of 
solar-type stars (e.g. Abt \& Levy 1976; Duquennoy \& Mayor 1991) found 
relatively high binary frequencies ($\ga$60\%) and a wide range of binary 
separations ($\la$$10^4$ AU) and mass ratios (1 to $\la$0.1). This has led to 
the common assumption that binary systems are the primary channel for star 
formation. However, multiplicity surveys of lower-mass M dwarfs (Fischer \& 
Marcy 1992; Reid \& Gizis 1997) observed marginally lower binary frequencies 
(35-43\%) and surveys near and below the substellar boundary (Close et al. 2003; 
Bouy et al. 2003; Burgasser et al. 2003;  Siegler et al. 2005) found 
substantially lower binary frequencies (10-20\%) and separations (typically 
$\la$20 AU) and a strong tendency toward mass ratios near unity.

These results demonstrate that field binary properties depend on mass. 
Unfortunately, binary properties for field stars are reported only for broad 
mass ranges, so they do not place strong constraints on the functional form of 
this dependence. Various groups interpret the transition in binary properties as 
either a sharp break near the stellar/substellar boundary (Kroupa et al. 2003;  
Close et al. 2003) or a smooth mass dependence (Luhman et al. 2004d). Also, 
field multiplicity surveys cannot constrain the mass dependence of substellar 
binary properties due to the degeneracy between brown dwarf masses and ages. 
Substellar companions in the field also tend to be old and intrinsically faint, 
so a limited range of binary mass ratios are accessible to observations. 
Finally, the field represents a composite population drawn from all 
star-formation regions, so field surveys cannot probe the dependence of binary 
properties on initial conditions (the stellar density, total mass, or mean Jeans 
mass of the formation region). One solution to these problems is to extend 
multiplicity surveys to the nearest young uniform stellar populations: OB 
associations, T associations, and open clusters.

Multiplicity surveys have been conducted for many of the bright members of 
nearby open clusters and associations over the past decade using HST (Martin et 
al. 2000; Luhman et al. 2005), adaptive optics (Patience et al. 2002; Bouy et 
al. 2006b), and speckle interferometry (Kohler et al. 2000; White et al.  
2006). These surveys have confirmed many trends observed in the field, such as 
the high binary frequency and separations of solar-type stars (e.g. Ghez et al. 
1993; Kohler et al. 2000) and the low frequency and separation of the 
lowest-mass systems (Martin et al.  2003; Luhman et al. 2005; Kraus et al. 2005, 
2006; Bouy et al. 2006a). However, they have also found some potentially 
interesting discrepancies. Surveys of different regions have revealed a mass 
dependence in binary frequency that is either smooth (Taurus-Auriga; White et 
al. 2006) or potentially discontinuous near the substellar boundary (Upper 
Scorpius; Kohler et al. 2000; Kraus et al. 2005). Several systems with unusually 
wide separations or low mass ratios have also been found (e.g. Luhman et al. 
2004b; Bouy et al. 2006b).

High-resolution imaging techniques are typically resource-intensive, so it is 
expensive to undertake large programs which can sample a wide range of mass with 
sufficient statistical significance to characterize these effects. However, wide 
binary systems in the nearest associations have angular separations large enough 
to resolve without these techniques. A program which exploits a uniform, 
high-quality seeing-limited survey could substantially enhance our understanding 
of the role of mass and environment in binary properties.

In this paper, we present the results from a search for new young binary systems
in the Two-Micron All-Sky Survey (2MASS), an all-sky imaging survey conducted in
the near-infrared. In Section 2, we describe the selection of our survey sample, 
and we describe our data processing techniques in Section 3. We summarize the 
results of our search in Section 4. Finally, in Section 5 we compare these 
results to the standard paradigm of stellar multiplicity and discuss the 
implications for the processes of multiple star formation.

\section{Sample Selection}

In Table 1, we describe the young associations from which we have drawn our
sample: Taurus-Auriga, Chamaeleon I, and the two proposed subgroups of Upper
Scorpius.  The sample regions have been selected to include all large stellar
populations ($\ga$100 known members) which are not heavily embedded, are
located at distances of $\la$200 pc, and have ages $\la$30 Myr. These criteria
neglect small associations and moving groups which can not contribute
significant statistics (TW Hya, MBM 12, Chamaeleon II, $\eta$ and $\epsilon$
Cham, and the Lupus clouds), distant populations for which seeing-limited
observations on a small telescope cannot probe sufficiently small separations
(IC 348 and the subgroups of Orion), embedded populations like the $\rho$ Oph
complex, and old populations in which the wide binary population may have been
shaped by dynamical evolution (Praesepe, Pleiades, $\alpha$ Persei).

In Table 2, we list the association members which we have adopted as our primary 
sample in the multiplicity search. The regional membership of our sample has been 
confirmed via low-resolution spectroscopy to verify signatures of youth, so 
contamination of the primary sample should be negligible. As we discuss in Section 
4.2, the surveys from which we draw our sample are likely to be incomplete due to 
selection biases; many of the new candidate companions found here would have been 
identified in previous surveys if they were complete and unbiased. This could 
potentially cause us to overestimate the wide binary frequency. Wide binaries 
would only be excluded from our sample if both components were absent from 
previous membership surveys, so they are less likely to have been omitted from our 
sample than single members. However, this effect would have been more prevalent 
among faint low-mass systems (where incompleteness is higher). We are testing for 
a decline in the binary frequency with mass, and any detection of this trend would 
be robust against this bias.

Saturation occurs for 2MASS sources brighter than $K\sim$8, but the images can 
still be used for sources as bright as $K\sim$6; we have neglected only the 
high-mass association members which are brighter than this limit, corresponding 
to spectral types earlier than G0. These bright stars typically have been 
studied with adaptive optics (e.g. Kouwenhoven et al. 2005 for the Sco-Cen 
complex), so analysis of 2MASS data would not contribute significant new 
results. We also omit all sources which do not have confirmed spectral types 
since we can not estimate their mass.  This criterion should eliminate most of 
the sources which are embedded in massive envelopes and surrounded by resolved 
nebulosity.  Finally, four of our primary sample members are fainter than 
our detection limit for binary companions ($K=14.3$), but we retain them 
in our sample in case they are binary companions to higher-mass 
association members which have not yet been identified.

In the following subsections, we briefly describe each association and summarize 
the construction of our search sample.

\begin{deluxetable}{lcclc}
\tabletypesize{\scriptsize}
\tablewidth{0pt}
\tablecaption{Nearby Young Associations}
\tablehead{\colhead{Name} & \colhead{Distance} & 
\colhead{Age} & \colhead{Type} & \colhead{Members} \\
\colhead{} & \colhead{(pc)} & \colhead{(Myr)}
}
\startdata
Chamaeleon I&170&1-2&T Assoc.&147\\
Taurus-Auriga&145&1-2&T Assoc.&235\\
Upper Scorpius A&145&5&OB Assoc.&356\\
Upper Scorpius B&145&5&OB Assoc.\tablenotemark{a}&45\\
\enddata
\tablenotetext{a}{As we discuss in Appendix C, the nature of Upper Sco B is 
still uncertain.}
\end{deluxetable}

\begin{deluxetable*}{llccrrrccrcl}
\tabletypesize{\scriptsize}
\tablewidth{0pt}
\tablecaption{Confirmed Members of Nearby Young Associations}
\tablehead{\colhead{Name} & \colhead{Region} & \colhead{RA} & 
\colhead{DEC} & \colhead{$K$} & \colhead{$J-K$} & \colhead{$H-K$} &
\colhead{SpT} & \colhead{Mass} & \colhead{$\chi$$_3$\tablenotemark{a}} &
\colhead{$\mu_{\alpha}$,$\mu_{\delta}$} & \colhead{References}
\\
\colhead{} & \colhead{} & \multicolumn{2}{c}{(Eq=2000)} & \colhead{}
& \colhead{} & \colhead{} & \colhead{} & \colhead{($M_{\sun}$)} & \colhead{} &
\colhead{(mas yr$^{-1}$)}
}
\startdata
ScoPMS005&UScoA&15 54 59.86&-23 47 18.2&7.03&0.54&0.16&G2&1.66&26.57&-28,-38&Walter et al. (1994)\\
ScoPMS013&UScoA&15 56 29.42&-23 48 19.8&8.75&0.92&0.23&M1.5&0.54&2.00&16,-42&Walter et al. (1994)\\
ScoPMS014&UScoA&15 56 54.97&-23 29 47.8&10.29&0.93&0.30&M3&0.36&1.67&-8,-28&Walter et al. (1994)\\
\enddata
\tablenotetext{a}{The $\chi$$_3$ statistic is a measure of how well each 
object is fit by a single point source; see Section 3.2.}
\tablecomments{The full table of 783 sample members can be accessed from
the website http://www.astro.caltech.edu/$\sim$alk/ (in PDF, TeX, or
plain-text formats) and will be posted as online-only content in ApJ.}
\end{deluxetable*}

\begin{deluxetable*}{lccccccccccc} 
\tabletypesize{\scriptsize}
\tablewidth{0pt} 
\tablecaption{Close Pairs of Confirmed Association Members\label{tbl3_01}} 
\tablehead{\colhead{Name} & \multicolumn{3}{c}{Primary} & 
\multicolumn{3}{c}{Secondary} & \colhead{Projected} & 
\colhead{Position}
\\
\colhead{} & \colhead{$J-K$} & \colhead{$H-K$} & \colhead{$K$} & 
\colhead{$J-K$} & \colhead{$H-K$} & \colhead{$K$} &
\colhead{Sep(\arcsec)} & \colhead{Angle(deg)} 
} 
\startdata 
DHTau&1.59&0.65&8.18&0.93&0.21&8.39&15.23&126\\
DITau\\
FVTau&2.48&0.88&7.44&1.93&0.62&8.87&12.29&105.7\\
FVTau/c\\
FZTau&2.55&1.05&7.35&1.93&0.62&8.05&17.17&250.5\\
FYTau\\
GGTau A&1.31&0.45&7.36&1.09&0.42&9.97&10.38&185.1\\
GGTau B\\
GHTau&1.32&0.44&7.79&1.19&0.40&6.96&21.77&15.2\\
V807Tau\\
\enddata 
\tablecomments{The full version of this table has been moved to the end of this
manuscript (Table 8) to enhance readability.}
\end{deluxetable*}

\subsection{Scorpius-Centaurus}

The Sco-Cen OB Association consists of three distinct subgroups: Upper Scorpius 
(USco; 5 Myr and 145 pc), Upper Centaurus-Lupus (UCL; 13 Myr and 160 pc), and 
Lower Centaurus-Crux (LCC; 10 Myr and 118 pc) (de Geus et al. 1989; de Zeeuw et 
al. 1999). Sco-Cen has been recognized for nearly a century as a moving group of 
early-type stars (e.g. Kapteyn 1914; Blaauw 1946;  Bertiau 1958; Jones 1971). 
However, surveys to identify low-mass stellar members have been undertaken only 
in the past 15 years and have concentrated almost exclusively on USco. Initial 
surveys (Walter et al. 1994;  Kunkel 1999)  identified candidate members from 
surveys for X-ray emission, while subsequent surveys (Preibisch et al. 1998; 
Ardila et al. 2000; Preibisch et al.  2001; Preibisch et al. 2002; Martin et al. 
2004; Slesnick et al. 2006a) used wide-field optical/NIR surveys to select 
candidate members with colors and magnitudes consistent with the assumed age and 
distance. Membership was confirmed with low- or intermediate-resolution 
spectroscopy to confirm indicators of youth such as lithium absorption, 
H$\alpha$ emission, or low surface gravity.  Proper-motion member identification 
is typically not possible for faint Sco-Cen members since their proper motions 
are not sufficiently distinct from those of background stars; the only major 
effort has been by Mamajek et al.  (2003), who identified candidate G- and 
K-dwarf members of UCL and LCC based on proper motions, then confirmed their 
membership with low-resolution spectroscopy.

The sample sizes for UCL and LCC are marginal ($\sim$50 members each) and span
a limited range of masses, and the associations' low galactic latitude
($|b|<20$) results in substantial contamination from reddened background
stars, so we have chosen to only consider Upper Sco. We select our sample from
the surveys of Walter et al. (1994), Preibisch et al. (1998), Kunkel (1999),
Ardila et al. (2000), Preibisch et al. (2001), Preibisch et al. (2002), Martin
et al. (2004), and Slesnick et al. (2006a).

Brandner et al. (1996) noted that some of the objects in these surveys form a
distinct subgroup in the southwest, near the border with UCL;  they named the
main population Upper Sco A and the subgroup Upper Sco B (hereafter USco-A and
USco-B).  A multiplicity survey by Kohler et al. (2000) subsequently found that
these two populations might have distinct binary statistics, with a much wider
mean separation in USco-B. As we show in Appendix B, the members of USco-B also
appear to have distinct kinematics. These results suggest that USco-B should be
treated as a distinct population. Based on the population kinematics and the
previous dividing lines adopted by Brandner et al. and Kohler et al., we assign
all sample members west of 16h and south of -28 deg to USco-B, and all remaining
members to USco-A. It is quite likely that there is some overlap along this
border, but the precision of the kinematic data does not allow us to
unambiguously determine this or to establish the subgroup membership of 
individual sources.

We also note that two USco members, ScoPMS008A and ScoPMS008B, are located 
$\sim$15\arcsec\, from an early type USco member, HD 142424 (A8IV/V; de Zeeuw et 
al. 1999). It is possible that these stars are companions to HD 142424 and not 
independent primaries; since they fall within our identification range for binary 
companions in USco-A ($\la$20\arcsec), then we do not treat these sources as 
independent primaries. Kohler et al. (2000) found (and we verify) that ScoPMS008A 
is itself a binary system with a separation of $\sim$1.6\arcsec, which suggests 
that this could be at least a quadruple system.

\subsection{Taurus}

The Taurus-Auriga association (Taurus; 1-2 Myr; 145 pc; Bertout et al. 1999; 
White \& Ghez 2001) has been recognized for more than 60 years as the nearest 
northern site of low-mass star formation and is the home of the archetypical 
star T Tauri. The low-mass stellar population of Taurus-Auriga has been 
classified gradually over the this time period (e.g. Joy et al. 1945;  Herbig et 
al. 1952; Cohen \& Kuhi 1979); unlike Sco-Cen, Taurus is largely devoid of stars 
more massive than 1-2 $M_\sun$.

A census of known Taurus members was presented in Kenyon \& Hartmann (1995) and 
has been supplemented by additional surveys to identify very-low-mass stellar 
and substellar members of Taurus-Auriga by Briceno et al. (1993, 1998, 2002), 
Strom \& Strom (1994), Martin et al. (2001), Luhman et al. (2003a, 2004d, 2006), 
Guieu et al. (2005), and Slesnick et al. (2006b). Members of Taurus-Auriga have 
also been confirmed in a followup survey of continuum (heavily veiled) sources 
by White \& Basri (2003) and a survey for Hyades members by Reid \& Hawley 
(1999). Finally, it was pointed out by White et al. (2006) that the source FV 
Tau/c2 (Hartigan et al. 1994) was omitted from the compilation of Kenyon \& 
Hartmann. We have constructed our Taurus source list from the Kenyon \& Hartmann 
census, plus all subsequent surveys.

\subsection{Chamaeleon I}

The Chamaeleon I complex (ChamI; 1-2 Myr; $\sim$160-170 pc; Whittet et al. 1997;  
Wichmann et al. 1998; Bertout et al. 1999) is another nearby site of ongoing
star formation. Like Taurus-Auriga, it is composed primarily of low-mass stars
and molecular clouds and possesses few high-mass stars. Much of its stellar
population was identified by optical and near-infrared surveys during the 1970s
and 1980s (e.g Henize \& Mendoza 1973; Schwartz 1977; Glass 1979; Baud et al.
1984).

Carpenter et al. (2002) and Luhman (2004b) have compiled censuses of known 
members and candidate members based on these and other surveys, and Luhman 
(2004b) confirmed the membership of many candidate members using optical and NIR 
spectroscopy. An objective prism survey of the entire cloud by Cameron et al. 
(2004) also confirmed 4 additional candidate members and identified 7 new 
members. Finally, one candidate substellar member from the survey of Oasa et al. 
(1999) was spectroscopically confirmed as a ChamI member by Luhman et al. 
(2004c). We have constructed our ChamI sample from the 151 confirmed ChamI 
members of Luhman (2004b), Luhman et al. (2004c), and Cameron et al. (2004) with 
spectral types later than G0.

\subsection{Spectroscopically Confirmed Stellar Pairs}

Spectroscopic surveys of these stellar associations have identified many close 
($<$30\arcsec)  pairs of members. Given the typical low surface density of 
association members on the sky, these stars could be gravitationally bound 
binary companions. We list these candidate binary systems in Table 3. Many 
systems have projected separations lower than our survey's outer identification 
limits (Section 4.1); in these cases, we have removed the secondary star in each 
pair from our statistical sample. Candidate secondaries at wider separations are 
considered to be independent systems for statistical purposes.

\section{Observations and Data Analysis}

\subsection{The Two-Micron All-Sky Survey}

The Two-Micron All-Sky Survey (2MASS; Skrutskie et al. 2006) observed 99.998\% of
the sky in the $J$, $H$, and $K_s$ bands over an interval of 4 years. Each point
on the sky was imaged six times and the coadded total integration time was 7.8s,
yielding 10$\sigma$ detection limits of $K=14.3$, $H=15.1$, and $J=15.8$. The
saturation levels depend on the seeing and sky background for each image, but 
are typically $J<9$, $H<8.5$, and $K_{s}<8$. The pixel scale of the detector was 
2 arcsec pix$^{-1}$, but acquisition of multiple images allowed for subsampling to 
increase the effective resolution; the final pixel scale for each processed 
image is 1 arcsec pix$^{-1}$, which critically samples stellar point-spread functions 
(PSFs) given a typical resolution of 3$\arcsec$ FWHM. The typical astrometric 
accuracy attained for the brightest unsaturated sources ($K\sim8$) is $\sim$100 
mas, and the photometric zero-points are calibrated to $<$0.02 mag.

The 2MASS Point Source Catalog (PSC; Cutri et al. 2003) and the processed survey 
images are available from the 2MASS 
website\footnote{http://www.ipac.caltech.edu/2mass/}.  We use PSC data to 
identify all wide ($>$5\arcsec) visual companions to our sample members. 
However, the PSC does not always distinguish multiple point sources in close 
proximity ($\la$5\arcsec), instead reporting only the brightest source.  This 
suggests that wide neighbors to our sample members should be identified in the 
PSC, but most close neighbors are probably absent.

We address this incompleteness by working directly with the processed survey 
images to identify close ($\la$5\arcsec) companions via PSF-fitting photometry.  
From the 2MASS website, we extracted postage-stamp (60x60$\arcsec$) and 
wide-field (510x1024$\arcsec$) images for each of the association members 
described in Section 2. The wide-field images were used to create reference PSFs 
for each science target, while the postage-stamp images have been used to 
identify close visual companions. The width of the wide-field images 
(510$\arcsec$) corresponds to the width of each 2MASS survey tile; any image 
with larger width would include data taken at different epochs, and therefore 
with different seeing conditions. The height was chosen to allow for $\ga$10 PSF 
reference stars brighter than $K\sim$11 in all fields. The size of the overlap 
region between adjacent tiles was 60\arcsec\, in right ascension and 
8.5\arcmin\, in declination, so each science target appeared to be $>$30\arcsec 
away from the edge in at least one tile.

The 2MASS survey images were produced by coadding multiple exposures taken in 
sequence, each offset by $\sim$85\arcsec\, in declination, so drawing PSF 
reference stars from several arcminutes away could lead to nonuniform images. 
Only sources $\la$40$\arcsec$\, north or south of a science target were observed 
in all six exposures that the science target was observed, and sources 
$\ga$500$\arcsec$\, north or south do not share any simultaneous scans. However, 
all of the scans which contribute to a wide-field image were observed within 
$\sim$30 seconds. We do not expect the seeing-based PSF to change on this short 
timescale, and we have found that the PSF is usually constant over each entire 
wide-field image ($\sigma$$_{FWHM}$$\sim$0.1$\arcsec$).

\subsection{Data Reduction and Source Identification}

We identified candidate companions and measured their fluxes from the 
postage-stamp image of each sample member using the 
IRAF\footnote{IRAF is distributed by the National Optical 
Astronomy Observatories, which are operated by the Association of Universities 
for Research in Astronomy, Inc., under cooperative agreement with the National 
Science Foundation.} package DAOPHOT (Stetson 1987), specifically with the 
PSF-fitting photometry routine ALLSTAR. The template PSFs for each 
postage-stamp image were created 
using the PSTSELECT and PSF tasks. We selected template stars for each source 
from the corresponding wide-field image; each PSF was based on the eight 
brightest, unsaturated stars which appeared to be isolated under visual 
inspection. The appropriate photometric zero-point was extracted from the image 
headers. We compared PSF-fitting magnitudes for single stars to the 
corresponding PSC values in order to test our results; there is no systematic 
offset, and the standard deviation of the random scatter in $m_{PSF}-m_{PSC}$ is 
$\sim$0.03 magnitudes.

As we have discussed in previous publications (Kraus et al. 2005, 2006), one 
limitation of ALLSTAR-based PSF photometry is that binaries with very close 
($\la$$\theta$$_{FWHM}$) separations are often not identified, even when their 
combined PSF deviates significantly from that of a true point source. This 
limitation can be overcome for known or suspected binaries by manually adding a 
second point source in approximately the correct location and letting ALLSTAR 
recenter it to optimize the fit. However, this method requires objective 
criteria for identifying suspected binaries;  subjective selection methods like 
visual inspection would not allow us to rigorously choose and characterize a 
statistically complete sample. We have found that ALLSTAR's $\chi$$^2$ 
statistic, which reports the goodness-of-fit between a source and the template 
PSF, is an excellent diagnostic for this purpose. Since there are images in 
three bandpasses, we use a single diagnostic value, denoted $\chi$$_3$, which is 
the sum of the three $\chi$$^2$ values obtained for each association member when 
fit with a single point source. We list the value of $\chi$$_3$ for each 
association member in Table 2.

In Figure 1, we plot the values of $\chi$$_3$ as a function of K-band
magnitude for a subset of sample members with no known companions between
0.5$\arcsec$ and 15$\arcsec$ (according to the surveys of Leinert et al.
1993; Ghez et al.  1993;  Simon et al. 1995; Duchene 1999; Kohler et al.
2000; Kraus et al.  2005, 2006; White et al. 2006). The goodness of fit
degrades rapidly for saturated stars ($K\la8$), so our technique does not
discriminate betweeen single stars and candidate binaries in this regime.
However, since there are few stars brighter than the saturation limit, we
decided not to reject them until we were certain we could not identify any
binary systems via other methods. The distribution of $\chi$$_3$ values for
unambiguously unsaturated stars ($K>8.5$) is not normally distributed, but
95\% of these stars produce fits with $\chi$$_3$$<$2.5, so we have selected all 
sources with $\chi$$_3$$\ge$2.5 as candidate binary systems.

The mean value of $\chi$$_3$ for single stars should be $\sim$3 since it 
represents the sum of 3 variables which follow a $\chi$$^2$ distribution. However, 
we find that the mean value reported by ALLSTAR for unsaturated single sources is 
$\sim$1.75. This disagreement is caused by an overestimate of the photometric 
errors in each observation by ALLSTAR. The coadding and subsampling process used 
in the 2MASS image processing pipeline results in correlated noise between 
adjacent pixels of the final survey images, so the true uncertainties are lower 
than those estimated solely by Poisson statistics (Skrutskie et al. 2006).

We identified the candidate binaries in our sample based on this empirically 
motivated $\chi$$_3$ selection criterion, and then we attempted to fit each with 
a pair of point sources separated initially by the PSF FWHM (3$\arcsec$) and 
with position angle corresponding to the angle of maximum elongation of the 
system PSF. The ALLSTAR routine optimized the components' separation, position 
angle, and magnitudes to produce the optimal fit; as we further summarize in 
Section 3.3, known binaries were typically fit with consistent positions and 
flux ratios in all three bandpasses while contaminants (such as sources with an 
erroneous template PSF in one filter)  did not produce consistent fits in 
multiple images. We adopt the criterion that any candidate binary with component 
positions within 1$\arcsec$ (3$\sigma$ for astrometry of very close, faint 
companions; Section 3.4) in all three filters is a bona-fide visual binary. We 
found that saturated stars produced fits for erroneous companions at separations 
of 1.0-1.5\arcsec, so we have rejected all candidate companions to saturated 
targets ($K_{tot}<8$) with separations of $<$2\arcsec. Known binaries with wider 
separations produced consistent fits even in the saturated regime for systems 
fainter than $K\sim6$, so we adopted this as a maximum brightness limit for our 
sample.

Finally, we compared the location of each candidate companion with the online 
catalog of 2MASS image artifacts. We found that a candidate companion to 
MHO-Tau-4 was coincident with a persistence artifact flag. Furthermore, a 
previous high-resolution imaging survey with HST (Kraus et al. 2006) found no 
optical counterpart to a limit of $z'$$\sim$24, so we removed this candidate 
companion from further consideration and treat MHO-Tau-4 as a single star.

\begin{figure*}
\plotone{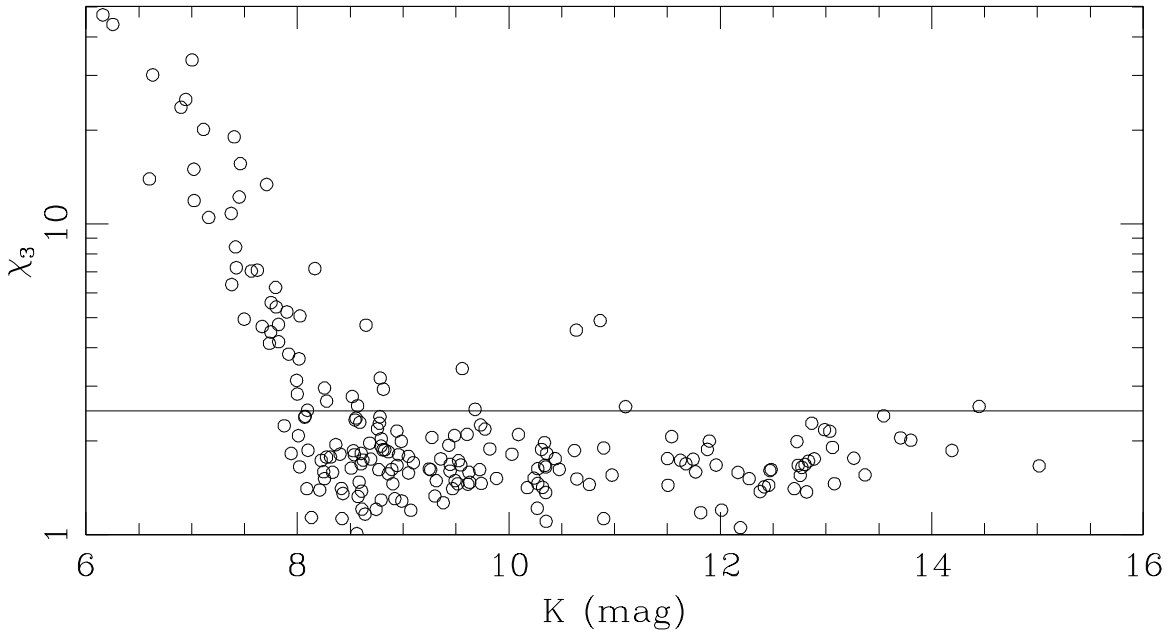}
\caption{A plot of the goodness-of-fit as a function of K-band magnitude for 203 
objects with no wide companions (0.5-15$\arcsec$). The sharp increase in 
$\chi$$_3$ at $K\sim$8 is due to the onset of image saturation; the stars in 
this brightness range are typically late K or early M, so saturation begins 
simultaneously in all three bands. The solid line at $\chi$$_3$$=2.5$ denotes 
the 95\% confidence interval for nominally single stars; we have selected all 
sample members above this limit as candidate close binaries. We found that our 
fitting algorithm for identifying companions is effective for mildly saturated 
stars, so we include association members up to $K=6$.
} 
\end{figure*}

\subsection{Sensitivity Limits}

We determined companion detection limits as a function of distance from the 
primary stars via a Monte Carlo simulation similar to that of Metchev et al. 
(2003).  We used the IRAF task DAOPHOT/ADDSTAR to add artificial stars at a 
range of radial separations and magnitudes to the images of FO Tau, MHO-Tau-5, 
KPNO-Tau-8, and KPNO-Tau-9. These four sources have been shown to be single to 
the limits of high-resolution imaging (Ghez et al. 1993; Kraus et al. 2006) 
and span the full range of brightness in this sample. We then attempted to 
identify the artificial companions via PSF-fitting photometry. Our photometric 
routines attempt simultaneous source identification in all three filters in 
order to separate erroneous detections from genuine companions, so we created 
the same synthetic source in all three filters using colors from the 2 Myr 
Baraffe isochrones (Baraffe et al. 1998).

In Figure 2, we show our survey's 50\% detection limits as a function of 
separation for identifying candidate companions using the same PSF-fitting 
algorithm as our actual search program. The minimum separation at which we can 
detect equal-flux companions is $\sim$1\arcsec\, for bright, unsaturated sources 
and $\sim$1.6\arcsec\, for sources just above our adopted $K$ band magnitude 
limit ($K=14.3$). The 10\% and 90\% detection limits are typically $\sim$0.5 
magnitudes below and above the 50\% limit. The sensitivity of PSF-fitting 
photometry falls at separations $\ga$5\arcsec\, since objects become cleanly 
resolved and most companion flux falls outside the fitting radius for the 
primary. However, the PSC is complete to at least $K=14.3$ at larger separations, 
so wider companions will be recovered by our search of the catalog.

We also show the separation and flux ratio for known binary systems which 
have been detected in K-band surveys (Kohler et al.  2000; White et al. 
2006) and whether these systems were unambiguously recovered (via either 
PSF-fitting photometry or the PSC), identified as candidate systems based on 
the $\chi$$_3$ criterion, or not recovered. The limits between detected and 
nondetected systems are roughly consistent with our empirically determined 
magnitude limits, but there are few known systems which fall near these 
limits. There are only two known wide systems among the faintest members of 
our sample ($K>11$), so we can not significantly test the detection limits 
of our search method in this brightness range. However, we identified four 
additional candidate companions to sources in this brightness range, plus 
numerous likely background stars, so our survey appears to be sensitive to 
companions in this regime.

\begin{figure} 
\plotone{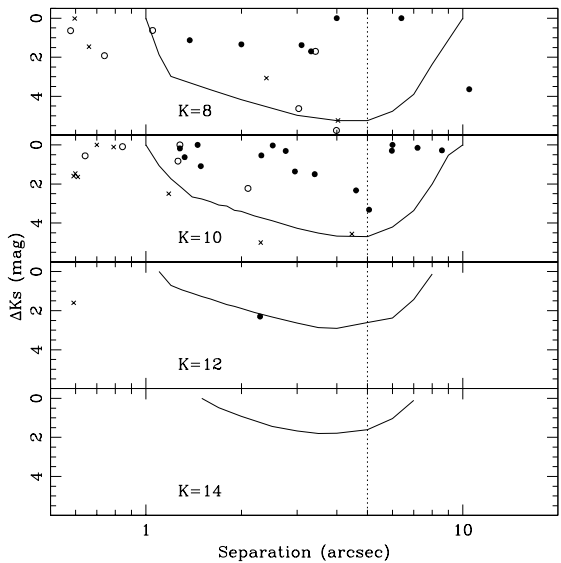} 
\caption{Detection frequencies as a function of separation for 
artificially-introduced companions to four known single objects spanning the 
survey sample's brightness range: FO Tau ($K=8.12$), MHO-Tau-5 ($K=10.06$), 
KPNO-Tau-8 ($K=11.99$) and KPNO-Tau-9 ($K=14.19$). The solid lines denote the 
50\% detection limit for our PSF-fitting photometry. The symbols represent 
known binary companions from high-resolution K-band multiplicity surveys in 
Upper Scorpius (Kohler et al. 2000) and Taurus (White et al. 2006 and 
references therein). Filled circles denote companions which we recovered, open 
circles denote companions which passed our $\chi$$^2$ criterion but did not 
produce significant fits, and crosses denote companions which were not 
recovered. The dotted line shows the minimum separation at which the PSC will 
identify all companions bright enough to be considered in our search 
($K<14.3$).
} 
\end{figure}

\begin{figure} 
\plotone{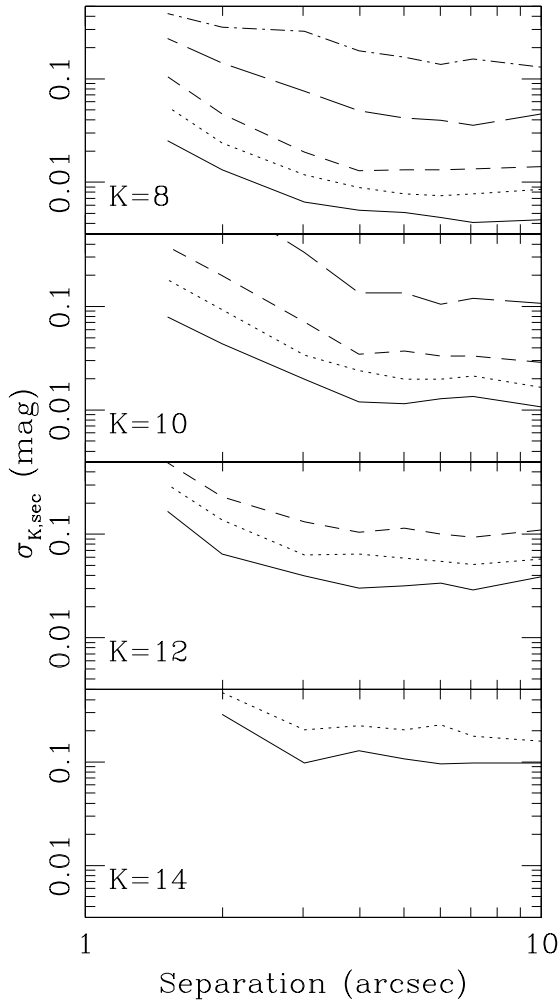} 
\caption{
The uncertainty in the measured binary companion brightness as a function of 
separation for simulated binary images spanning the range of primary and 
secondary brightnesses. The flux ratios shown are $\Delta$$K=$0, 1, 2, 4, and 6 
(solid, dotted, short-dashed, long-dashed, dash-dotted lines, respectively). The 
photometric uncertainties increase sharply at separations of $\la$3\arcsec, 
suggesting that observed photometric colors will not be accurate in 
this separation range.
} 
\end{figure}

\begin{figure} 
\plotone{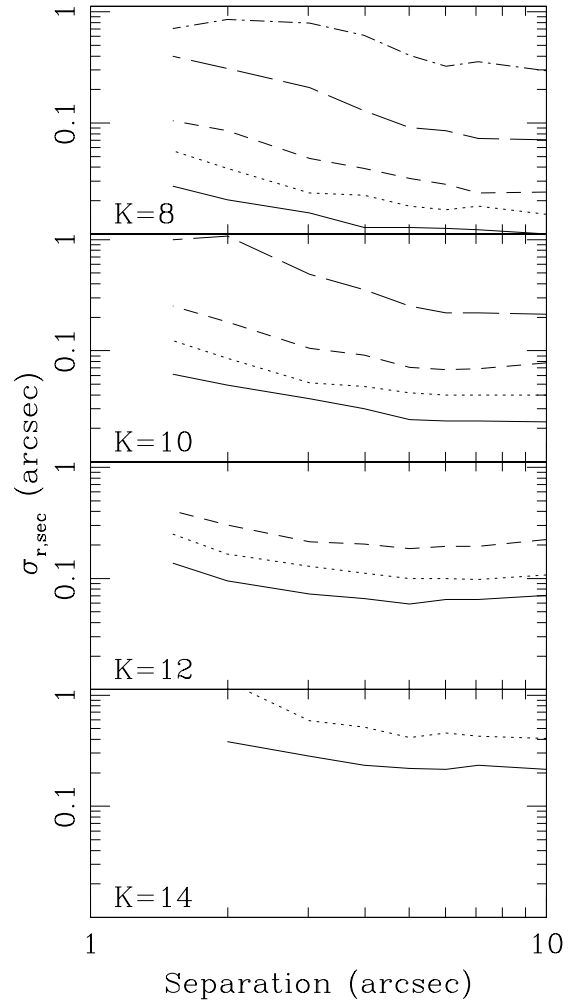} 
\caption{
As in Figure 3, showing uncertainties in binary secondary positions as a 
function of separation. 
} 
\end{figure}

\subsection{Uncertainties in Binary Properties}

Many of our candidate binaries have separations of $\la$$\theta$$_{FWHM}$, 
so our measurements could be subject to significant uncertainties. We 
tested these uncertainties by using a Monte Carlo routine to produce 
synthetic images for binaries spanning a range of primary brightnesses, 
flux ratios, and separations. Specifically, we used ADDSTAR to construct 
simulated $JHK$ images, and then we measured the binary fluxes and 
separations for each set of simulated images using ALLSTAR. For each 
combination of parameters, we produced 100 sets of synthetic images with 
randomly distributed position angles. The $J-K$ and $H-K$ colors for the 
secondaries were drawn from the 2 Myr isochrone of Baraffe et al. (1998) 
in order to determine realistic values for $\Delta$$K$, $\Delta$$H$, and 
$\Delta$$J$.

In Figure 3, we show the standard deviation in the measured brightness for our 
simulated binary companions as a function of separation. These simulations 
predict that photometric uncertainties increase significantly at separations of 
$\la$3\arcsec, so measured colors may not be reliable at small separations. As 
we describe in Section 3.5, these colors are necessary at large separations 
($>$5\arcsec) to distinguish candidate companions from background stars. 
However, contamination from background sources should be low at small 
separations ($\la$3\arcsec) due to their low surface density, so we can neglect 
these selection criteria with only a minor increase in the number of erroneous 
binary identifications.

In Figure 4, we show a similar plot of the RMS scatter in the measured position 
of the secondary. The typical standard deviations are $\la$0.3\arcsec\, for all 
but the faintest companions, so the uncertainties in our measured separations 
should have similar precision. Given these positional uncertainties, the 
corresponding uncertainties in position angles range from 1 to 10 degrees, 
depending on the binary separation. The standard deviations in secondary 
position for our simulated images are consistent with the scatter between the 
three filters for each observed binary, so we adopt the results from these 
simulations as our estimated uncertainties.

We also conducted Monte Carlo tests to determine the probability of mistakenly
identifying a true single star as a binary. We constructed a series of simulated
images (100 each for four objects spanning our sample's range of brightness),
and then tried to fit each object with two point sources. We found that this
never produced consistent fits in 3 filters, though faint peaks due to noise
were occasionally identified in one of the 3 images. This suggest that the
probability of an erroneous binary identification due to statistical errors is
low ($<1\%$). This agrees with our results for known single stars; as we note in
Section 3.2, 5\% of known single stars fall above our $\chi$$_3$ criterion for
identifying candidate binaries. However, none of these yielded fits for multiple 
point sources in all 3 filters.

\subsection{Field Star Contamination}

The identification of binary companions based solely on proximity is
complicated by contamination from foreground dwarfs, background giants, and
reddened early-type background dwarfs. We have not conducted followup
spectroscopic or astrometric observations to confirm association membership,
so we must limit the survey to a total area in which the contamination from
background stars is small compared to the number of candidate binary
companions. We estimate the surface density of contaminants for each
association based on the total number of objects within an annulus of
30-90\arcsec\, from all of the association members in our sample. Field
surveys (e.g. Duquennoy \& Mayor 1991; Reid \& Gizis 1997) have identified few
binaries with projected separations of $\ga$500 AU ($\ga$30\arcsec at the
distance of our sample members), so this method will also address the
probability of chance alignment with other association members.

Our estimate of the contamination could be influenced by variations in
background source counts due to the large angular extent of these associations
or by variations in galactic latitude or extinction. The result would be a
systematic overestimation of the association probability for candidate
companions at points of high contamination and a corresponding underestimation
at points of low contamination. However, any local deviation from the mean
contamination rate should not affect the binary statistics for the association
as a whole since the ensemble background at 30-90\arcsec\, will match the
ensemble background at $<$30\arcsec. Our subsequent cuts against color, mass
ratio, and separation will also help to homogenize the sample by preferentially
removing background stars.

Most previous multiplicity surveys were based on observations in a single 
optical or near-infrared bandpass (e.g. Kohler et al. 2000); in the absence 
of color information, these surveys can only estimate physical association 
probabilities for candidate companions based on the surface density of 
background stars of similar brightness. Since 2MASS includes images in 3 
filters, we can reject most background stars by requiring colors consistent 
with regional membership (Section 4.1).  Specifically, we have plotted 
($K,J-K$) and ($K,H-K$) color-magnitude diagrams for each region and we 
require prospective binary companions to fall above a smoothed field main 
sequence (Bessell \& Brett 1988; Leggett et al. 2001)  for the regional 
distance in both CMDs. We have chosen to use $K$ as a proxy for luminosity 
instead of $J$ in order to minimize the effect of extinction for background 
stars. This choice will cause disk-bearing association members to sit 
preferentially higher in our color-magnitude diagrams, but this moves them 
further from our selection cutoff, so our results should be robust.

As a test of these color criteria, we have plotted color-magnitude diagrams for 
the members of our primary sample. We find that $\sim$97\% of the primaries have 
colors consistent with our definition of association membership, so any 
incompleteness in the selection of binary companions should be negligible. Most 
unselected primaries fall just below our color cuts; the only sample members 
which fall well below the association sequences are GSC 06191-00552 and 
USco-160803.6-181237.  Both of these objects are claimed to be 
spectroscopically-confirmed members of USco-A, but the spectra are not available 
in the literature. We have not detected any binary companions to these objects, 
so their erroneous inclusion in our sample would not significantly change our 
results. However, it might be prudent to reconsider their membership status with 
additional spectroscopic observations in the future.

\begin{deluxetable*}{lllllllllllll}
\tabletypesize{\scriptsize}
\tablewidth{0pt}
\tablecaption{Association Star Counts\label{tbl2}}
\tablehead{\colhead{} & 
\multicolumn{3}{c}{Chamaeleon I ($N=147$\tablenotemark{a})} & 
\multicolumn{3}{c}{Taurus-Auriga ($N=235$\tablenotemark{a})} & 
\multicolumn{3}{c}{USco-A ($N=356$\tablenotemark{a})} & 
\multicolumn{3}{c}{USco-B ($N=45$\tablenotemark{a})}
\\
\colhead{Sep} 
& \colhead{Source} & \colhead{Color} &\colhead{Bkgd}
& \colhead{Source} & \colhead{Color} &\colhead{Bkgd}
& \colhead{Source} & \colhead{Color} &\colhead{Bkgd}
& \colhead{Source} & \colhead{Color} &\colhead{Bkgd}
\\
\colhead{(\arcsec)} 
& \colhead{Count\tablenotemark{b}} & \colhead{Valid\tablenotemark{b}} 
&\colhead{Stars\tablenotemark{b}}
& \colhead{Count} & \colhead{Valid} &\colhead{Stars}
& \colhead{Count} & \colhead{Valid} &\colhead{Stars}
& \colhead{Count} & \colhead{Valid} &\colhead{Stars}
}
\startdata
0-3\tablenotemark{c}&7&-&0.9&9&-&0.9&15&-&2.0&8&-&0.3\\
3-5&5&5&1.0&6&5&0.7&8&4&0.3&1&0&0.1\\
5-10&8&6&4.8&10&5&3.1&12&5&1.4&6&4&0.4\\
10-15&19&12&8.0&22&11&5.2&32&8&2.4&3&0&0.6\\
15-20&20&13&11.2&23&13&7.2&36&6&3.4&4&0&0.8\\
20-25&34&18&14.4&21&12&9.3&44&6&4.3&5&1&1.1\\
25-30&39&28&17.6&33&16&11.4&60&5&5.3&9&4&1.3\\
30-90&766&461&-&733&298&-&1566&138&-&215&34&-\\
\enddata
\tablenotetext{a}{The total sample size for each region, as 
summarized in Table 1.}
\tablenotetext{b}{The number of unassociated contaminants was estimated 
from the surface density of sources which meet our color selection 
criteria in the 30-90\arcsec\, separation range; most of these sources 
should be foreground stars, background stars, or unbound association 
members.}
\tablenotetext{c}{We cannot use color criteria at separations of 
$<3$$\arcsec$\, due to the poor photometric precision (Section 3.4), so 
the surface density of unassociated contaminants is higher.}
\end{deluxetable*}

\begin{deluxetable*}{lccccccccccl} 
\tabletypesize{\tiny}
\tablewidth{0pt} 
\tablecaption{Candidate Wide Binary Systems\label{tbl3_02}} 
\tablehead{\colhead{Name} & \multicolumn{3}{c}{Primary} & 
\multicolumn{3}{c}{Secondary} & \colhead{Projected} & 
\colhead{Position} & \colhead{$\mu_{\alpha}$,$\mu_{\delta}$\tablenotemark{a}}
& \colhead{Ident} & \colhead{References}
\\
\colhead{} & \colhead{$J-K$} & \colhead{$H-K$} & \colhead{$K$} & 
\colhead{$J-K$} & \colhead{$H-K$} & \colhead{$K$} &
\colhead{Sep(\arcsec)} & \colhead{Angle(deg)} & \colhead{(mas yr$^{-1}$)} & 
\colhead{Method}
} 
\startdata 
2M11103-7722&2.00&0.68&10.03&2.21&0.77&13.85&9.30&108.8&-&PSC&-\\
C7-1&1.78&0.62&10.55&1.67&0.43&13.32&5.73&214.9&0,0&PSC&-\\
CHSM1715&2.05&0.85&10.90&1.42&0.43&13.94&9.07&30.3&-58,42&PSC&-\\
CHXR26&2.02&0.46&9.92&2.68&1.07&9.98&1.41&215.2&-&PSF&Luhman (2004b)\\
CHXR28&1.17&0.32&8.23&1.53&0.39&8.83&1.78&121.6&-&PSF&Brandner et al. (1996)\\
\enddata 
\tablecomments{The full table of sample members can be found in Table 9 at the 
end of this manuscript.}
\tablenotetext{a}{An entry of 0,0 denotes a source which was detected by 
the USNO-B survey, but did not show a significant proper motion. An entry of "-" 
denotes a source which was not detected by the USNO-B survey.}
\tablenotetext{b}{ScoPMS052 B is also known as GSC06209-01312; Martin et al. (1998) 
identified it as a WTTS.}
\end{deluxetable*}

\begin{deluxetable*}{lcccccccccl} 
\tabletypesize{\tiny}
\tablewidth{0pt} 
\tablecaption{Ultrawide Visual Companions\label{tbl3_03}} 
\tablehead{\colhead{Name} & \multicolumn{3}{c}{Primary} & 
\multicolumn{3}{c}{Secondary} & \colhead{Projected} & 
\colhead{Position} & \colhead{$\mu_{\alpha}$,$\mu_{\delta}$} &
\colhead{References}
\\
\colhead{} & \colhead{$J-K$} & \colhead{$H-K$} & \colhead{$K$} & 
\colhead{$J-K$} & \colhead{$H-K$} & \colhead{$K$} &
\colhead{Sep(\arcsec)} & \colhead{Angle(deg)} & \colhead{(mas yr$^{-1}$)}
} 
\startdata
C1-6&3.92&1.68&8.67&1.93&0.80&14.10&27.58&156.0&&OTS12(candidate; Oasa et al. 1999)\\
C1-6&3.92&1.68&8.67&2.26&0.70&13.75&24.51&123.8&-&OTS14(candidate; Oasa et al. 1999)\\
Cam2-19&2.40&0.74&10.25&2.36&0.72&13.45&23.13&107.6&-&-\\
Cam2-42&2.44&0.73&9.16&2.09&0.50&13.51&27.64&261.7&-&-\\
Cam2-42&2.44&0.73&9.16&2.11&0.65&14.14&28.18&180.4&-&-\\
\enddata 
\tablecomments{The full table of sample members can be found in Table 10 at the 
end of this manuscript.}
\tablenotetext{a}{The source [CCE98] 2-26 is an ultrawide neighbor of both ChaHa7 and CHXR76; its 
physical association, if any, is uncertain.}
\tablenotetext{b}{Haro 6-5 B is a known member of Taurus (Mundt et al. 1984), but was not included as 
part of our statistical sample because its spectral type is uncertain.}
\tablenotetext{c}{USco-160428.4-190441 B is also known as GSC06208-00611; Preibisch et al. (1998) 
identified it as a field star.}
\tablenotetext{d}{SCH16075850-20394890 B is also known as T64-2; The (1964) identified it as a strong 
H$\alpha$ emitter.}

\end{deluxetable*}

\begin{figure*} 
\plotone{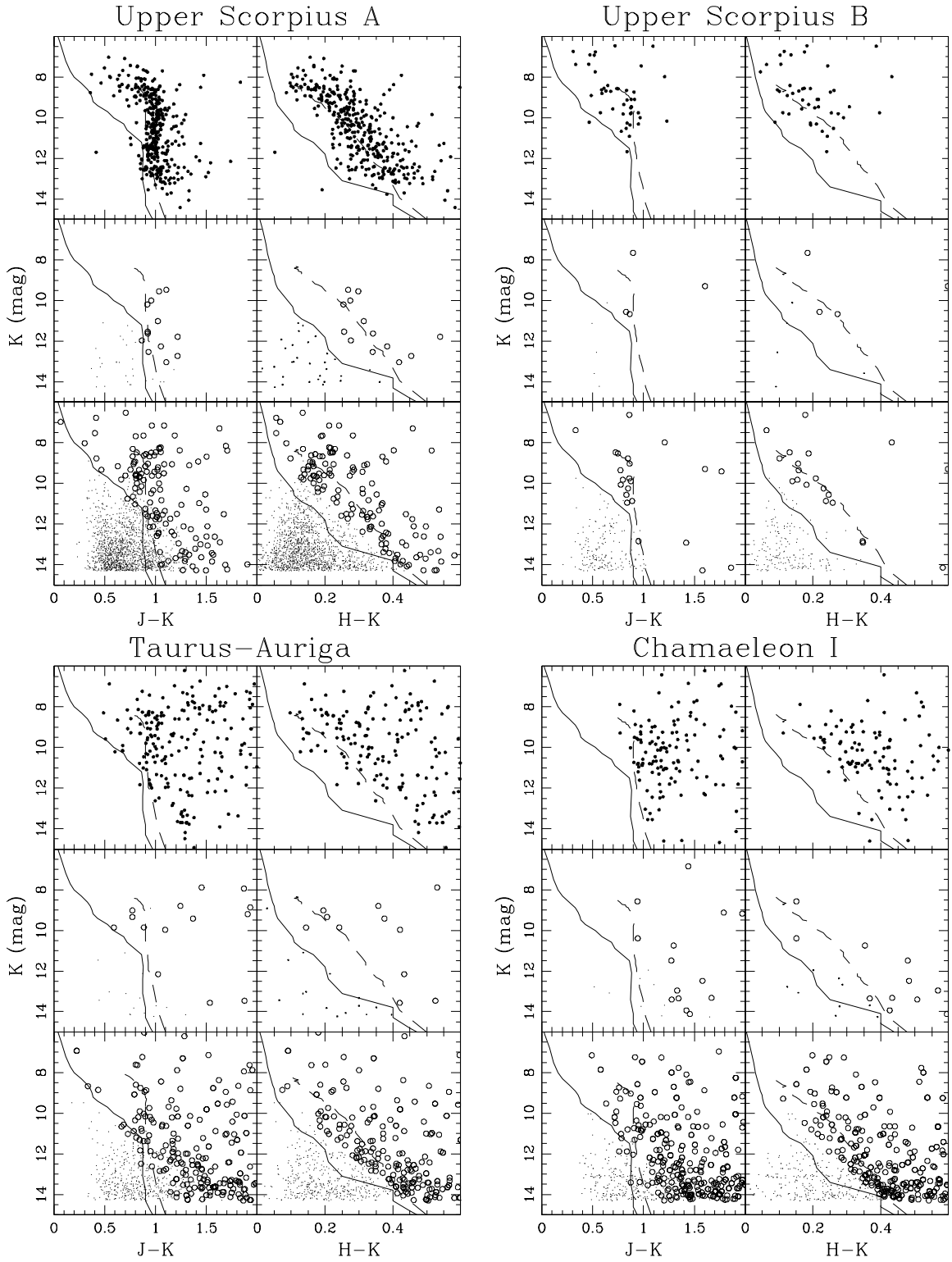} 
\caption{$K,J-K$ and $K,H-K$ color-magnitude diagrams for the four regions in our
survey.  The top panels show the confirmed association members in our
survey, the middle panels show all objects within 5-15\arcsec\, of known
association members, and the bottom panels show all objects within the
background annuli (30-90\arcsec). The solid line shows the main sequence at
the association distance and the dashed line shows the isochrone for the
adopted association age (Table 2). In the top panels, association members are 
shown with filled circles. In all other panels, sources which lie above a 
smoothed main sequence in both CMDs are shown with open circles and other 
sources are shown with small dots.
} 
\end{figure*}

\section{Results}

\subsection{Candidate Binary Companions}

We identified a total of 451 well-resolved visual companions brighter than 
$K=14.3$ within 30\arcsec\, of our sample members in the 2MASS PSC (Section 3.1), 
as well as 48 close ($\la$5\arcsec) candidate companions based on our PSF-fitting 
photometry of 2MASS image data (Section 3.2). We have chosen 30\arcsec\, 
($\sim$5000 AU) as an absolute upper limit for for identifying candidate 
companions since it corresponds to the maximum separation seen for field binaries 
at the distances of these association members. We also found 3280 visual 
companions within 30-90\arcsec\, of our sample members. Since the ratio of 
sources at 0-30\arcsec\, and 30-90\arcsec\, is roughly equal to the ratio of 
areas (1/8), we expect that most of the sources within 30\arcsec\, of our sample 
stars are foreground or background stars having colors inconsistent with 
association membership.

In Figure 5, we present ($K$,$J-K$) and ($K$,$H-K$) color-magnitude diagrams 
for the four regions showing all confirmed association members in our 
sample and all companions in two separation ranges (5-15\arcsec\, and 
30-90\arcsec) corresponding to likely companions and likely background stars. 
We summarize the number of objects which pass or fail the color selection 
criteria (Section 3.5) as a function of separation in Table 4. We also 
estimate the number of contaminants which are expected to pass both selection 
criteria in each separation range, assuming that the source density at 
30-90\arcsec\, represents the contaminant source density.

We showed in Section 3.4 that the uncertainties in our PSF-fitting photometry 
become significant at small separations, so we cannot use color criteria to 
identify candidate companions inside $\sim$3\arcsec. However, given the low 
surface density of background sources and the faintness of most nonmembers, we 
expect only a small level of contamination in this separation range. Each of the 
39 candidate companions at separations $<$3\arcsec\, has a sufficiently high 
probability of physical association ($\ga$80\%) to merit inclusion in our sample 
without using color cuts.

We have defined the maximum separation at which we identify candidate binary 
companions by requiring that the number of sources which pass our color 
selection requirement in each separation bin be $\ga$2 times the number of 
expected background companions. The corresponding probability that any 
individual source inside that separation limit is a background star will be 
$\la$50\%. Based on the expected contamination rates and visual companion 
counts in Table 4, these separations are 10\arcsec\, for ChamI, 15\arcsec\, for 
Taurus, 20\arcsec\, for USco-A, and 30\arcsec\, for USco-B. The separation limit 
is lower for regions with higher extinction since a higher fraction of 
background stars are reddened into our selection range. We adopt these 
separation limits as our criteria for identifying candidate binary companions. 
We note that sources at higher separations still have a nonnegligible 
probability of association, but the probability that any individual source is a 
binary companion will be low.

Using the color and separation cuts described above, we have identified (of 451 
sources identified in the PSC and 48 sources identified with PSF-fitting 
photometry) a total of 18 candidate binary companions in ChamI, 32 in Taurus, 40 
in USco-A, and 17 in USco-B. Of these candidates, 4, 7, 23, and 5, respectively, 
have not been previously reported in the literature.  We summarize the binary 
properties of these candidate systems in Table 5. Some of the very wide and very 
faint companions are likely to be unassociated foreground or background stars, so 
we will consider a restricted range of separations and mass ratios in our 
subsequent statistical analysis. In Table 6, we list the other visual companions 
with separations $<$30\arcsec\, (but wider than the association's companion 
identification limit) which have colors consistent with association membership and 
separations greater than the limits given above. Many of these sources are 
expected to be background stars, but additional information (such as optical 
photometry or kinematic data) could be used in the future to remove additional 
contaminants and more securely identify any ultrawide binary companions.

\subsection{Previous Observations}

Many of our candidate companions have been identified previously in the 
literature, but as we note in Tables 5 and 6, several of our candidates have also 
been rejected as association members based on the absence of spectroscopic 
signatures of youth. Some of the candidates we list have probably been considered 
and rejected in previous work, but most surveys do not publish their catalogue of 
confirmed field stars, so we cannot assess this number. 

We also find that five members of our sample (USco-160700.1-203309, 
SCH16151115-24201556, and USco80 in USco-A; 2MASSJ04080782+2807280 and 
2MASSJ04414489+2301513 in Taurus) have candidate companions which are 
significantly brighter, and thus are likely to be the system primary 
(making the known association member a binary secondary). This result is 
not surprising for the three Upper Sco members. Upper Sco is thought to 
contain several thousand low-mass members, and photometric surveys have 
identified many more candidates than could be confirmed via spectroscopy, 
so there are many more association members awaiting discovery. The two 
Taurus members are located on the edges of the association and were 
discovered by the only survey which considered these areas (Luhman 2006). 
Our newly-identified candidate companions are both brighter than the upper 
brightness limit for this survey ($H=10.75$), so there were no previous 
opportunities for them to have been discovered.

Finally, we find that 5 candidate companions identified in previous surveys have 
2MASS colors inconsistent with association membership: UX Tau B, V819 Tau B, 
HBC355 (HBC354 B), RXJ1524.2-3030B B, and RXJ1559.8-2556 B. Since $\sim$3\% of 
the spectroscopically confirmed association members in our primary star sample 
did not meet both color cuts, we expect (adopting the same percentage for the 
secondaries) that only $\sim$1-2 bona fide binary companions would not be 
selected. However, close pairs of stars have larger photometric errors, which 
increases the probability that some companions might fall outside our selection 
cuts. Of these five companions, three fall just below the color cuts (UX Tau B, 
HBC 355, and RXJ1524.2-3030B B) in our CMDs and the other two fall significantly 
below the color cuts, so we suggest that the first three are erroneous 
rejections, and therefore we keep these objects, while we consider the other two 
to be valid rejections.\footnote{V819 Tau B has also been classified as a 
background star by Woitas et al. (2001) due to its position on a (J,J-K) CMD and 
by Koenig et al. (2001) due to an absence of x-ray emission. UX Tau B and HBC355 
are spectroscopically confirmed cluster members, and no membership assessments 
are available for the other two sources.}

\subsection{Inferred Stellar Properties}

\begin{deluxetable*}{lccccccccl} 
\tabletypesize{\scriptsize}
\tablewidth{0pt} 
\tablecaption{Inferred Binary Properties\label{tbl3_04}} 
\tablehead{\colhead{Name} & \multicolumn{2}{c}{Primary}
& \multicolumn{2}{c}{Secondary} & \colhead{Projected} 
& \colhead{Mass}
\\
\colhead{}  & \colhead{SpT} & \colhead{Mass ($M_{\sun}$)}
& \colhead{SpT\tablenotemark{b} \tablenotemark{b}} 
& \colhead{Mass\tablenotemark{a} \tablenotemark{b} ($M_{\sun}$)} & 
\colhead{Separation(AU)\tablenotemark{b}} & 
\colhead{Ratio\tablenotemark{b} ($q$)}
} 
\startdata 
2M11103&M4&0.27&(M8.5)&(0.02)&1535&0.08\\
2M11103(/ISO250)&M4&0.27&M4.75(M5.5)&0.20(0.15)&1569&0.56\\
C7-1&M5&0.18&(M8)&(0.03)&945&0.18\\
CHSM1715&M4.25&0.25&(M7)&(0.05)&1497&0.18\\
CHXR26&M3.5&0.33&(M5)&(0.19)&233&0.57\\
\enddata 
\tablenotetext{a}{Values in parentheses are estimated from the system flux 
ratio $\Delta$$J$ and the spectroscopically determined properties of the 
primary.}
\tablenotetext{b}{Estimated statistical uncertainties are $\sim$10\% for 
mass ratios, $\sim$20\% for secondary masses, $\sim$2-3 subclasses for 
spectral types, and $\sim$10\% for projected separations.}
\tablecomments{The full table of sample members can be found in Table 11 at the 
end of this manuscript.}
\end{deluxetable*}

In Table 2, we list the inferred spectral types and masses for all of the 
association and cluster members in our sample. Spectral types are taken from the 
primary reference and were typically determined via low- or 
intermediate-resolution spectroscopy. We assume that the spectroscopically 
determined spectral type and mass for previously-unresolved binary systems 
corresponds to the primary mass and spectral type. Equal-mass binary components 
should have similar spectral types and the flux from inequal-mass systems should 
be dominated by the primary; in either case, spectroscopic observations of the 
unresolved system should have been affected only marginally by the flux from the 
secondary.

We estimated the masses of sample members by combining mass-temperature and 
temperature-SpT relations from the literature. No single set of relations spans 
the entire spectral type range of our sample, so we have chosen the M dwarf 
temperature scale of Luhman et al. (2003b), the early-type ($\le$M0) temperature 
scale of Schmidt-Kaler (1982), the high-mass stellar models of D'Antona \& 
Mazzitelli (1997; DM97), and the low-mass stellar models of Baraffe et al. (1998; 
NextGen). We apply the DM97 mass-temperature models for masses of $\ga$1 $M_\sun$ 
and the NextGen models for masses of $\la$0.5 $M_\sun$; in the 0.5-1.0 $M_\sun$ 
regime, we have adopted an average sequence. For each association, we adopt the 
models corresponding to the mean age listed in Table 1; this will introduce some 
uncertainty given the unknown age spread for each association. Large systematic 
errors may be present in these and all pre-main sequence models (e.g. Baraffe et 
al. 2002; Hillenbrand \& White 2004; Close et al. 2005; Reiners et al. 2005), so 
they are best used for relative comparison only.

Much of the uncertainty in theoretical mass-temperature relations can be assessed 
in terms of a zero-point shift in the mass; preliminary observational 
calibrations by the above authors suggest that theoretical models overestimate 
masses by 10-20\% over most of our sample mass range. This suggests that 
theoretical predictions of relative properties (e.g. mass ratios, 
$q=m_{s}/m_{p}$) might be more accurate than absolute properties (e.g. individual 
component masses) since the systematic mass overestimates will cancel. Relative 
quantities are also largely independent of age and extinction, which are expected 
to be similar for binary components. We have combined our adopted 
mass-luminosity-SpT relations with the near-infrared colors of Bessell \& Brett 
(1988) and the K-band bolometric corrections of Leggett et al. (1998, 2000, 2002) 
and Masana et al. (2006) to predict values for $q$ as a function of primary 
brightness $m$ and flux ratio $\Delta$$m$ in all three 2MASS filters. Some of our 
sample members could possess K-band excesses due to hot inner disks, so we have 
adopted the $q$ values predicted by the J-band fluxes; this will not eliminate 
the effect, but should minimize it. We have also combined our derived $q$ values 
with the estimated primary masses to predict secondary masses, and we use our 
mass-SpT relations to predict the corresponding secondary spectral types.

We list the derived values for each binary system in Table 7. Some wide 
binaries have independent SpT determinations for both components, so we 
report derived quantities with parentheses and measured quantities 
without. The typical uncertainties in $q$ are $\sim$10\% and represent the 
uncertainties in the photometry and the assigned spectral types, though 
some systematic effects (e.g.  unresolved multiplicity or different levels 
of extinction) could produce far larger values.  This can be seen in the 
discrepancies for some systems (e.g. GG Tau AB, MHO-2/1) which are known 
to be hierarchical multiple systems.  We can not quantify the unknown 
uncertainties in the theoretical models, but they should be considered 
when interpreting these results. The typical uncertainty in physical 
separation is $\sim$10\% and reflects the uncertainty in angular 
separation and the unknown depth of each system in its association; we 
assume each association has a total depth equal to its extent on the sky 
($\sim$40 pc for Taurus and Upper Sco, $\sim$20 pc for ChamI). The 
uncertainty in the mean association distance ($\sim$5 pc) introduces a 
systematic uncertainty of $\pm$3\%, but this is generally negligible.

\subsection{Binary Statistics}

Multiplicity surveys typically consider the frequency of binary systems for 
restricted ranges of parameter space (observed separations and mass ratios) 
corresponding to the survey completeness limits. For our analysis, we select a 
range of projected separations (330-1650 AU, set by the inner and outer detection 
limits of ChamI since those limits are most restrictive)  and flux ratios 
($\Delta$$K<2$, corresponding to $q\ga$0.25) that should be complete for all but 
the lowest-mass brown dwarfs in our sample. The inner separation limit and mass 
ratio limit are set by the resolution limit for low-mass sample members 
($K\sim$12.3) in ChamI, while the outer separation limit is set by the background 
contamination in ChamI, where our mass ratio cut allows us to choose a 90\% pure 
sample for separations $<$10\arcsec.

In Figure 6, we present plots of the wide binary frequency as a function of 
primary mass for each region in our sample. The binary fractions plotted 
correspond to our designated completeness regime: mass ratios $q>0.25$ and 
projected separations of 330-1650 AU. In the bottom panel, we show the field 
binary frequency in the same range of mass ratios and projected separations for 
solar-type stars (Duquennoy \& Mayor 1991), early-mid M dwarfs (M0-M6; Reid \& 
Gizis 1997), and brown dwarfs (Bouy et al. 2003; Burgasser et al. 2003). We also 
show the corresponding frequencies for early-type stars in USco-A and USco-B 
(Kouwenhoven et al. 2005). The bin sizes were chosen to evenly sample the mass 
range of our survey (0.025-2.50 $M_{\sun}$ for which the primary targets were 
brighter than our brightness cutoff ($K=14.3$). For each region in our survey, 
we also show the expected frequency for foreground and background sources which 
pass our color selection criteria and have $\Delta$$K<2$, assuming a background 
source count function N(K) matching that shown in Figure 2; in all cases, the 
expected contamination rate is negligible. USco-A, ChamI, and Taurus all show a 
decline in the binary frequency with mass, consistent with the results shown for 
field multiplicity surveys. USco-B does not show a decline, but the 
uncertainties are not small enough to strongly constrain the slope of any 
mass dependence.

This binary search may not be complete for objects in the lowest-mass bin where 
some binary companions could have been fainter than the survey detection limits 
($K>14.3$), so the true upper limits may be marginally higher. However, it has 
been observationally determined that most very low mass binaries in the field 
have mass ratios near unity ($q>0.7$) and much smaller separations ($\la$20 AU), 
so we are unlikely to have missed any wider or lower-mass ratio companions 
(Close et al. 2003; Burgasser et al. 2003; Bouy et al.  2003).

Another interesting distribution to consider would be the mass ratio
distribution for wide binaries as a function of mass and environment.
Unfortunately, extending our binary results along another axis of parameter
space exceeds the statistical limits of our sample, leaving most bins with
only 0-1 detections. The best solution for this is to combine all regions into
a single population. In Figure 7, we plot the mass ratio distribution in our
survey separation range (330-1650 AU) for the three highest-mass bins. We also
show the best-fit distribution for solar-type stars in the field (Duquennoy \&
Mayor 1991).

This result should be treated with caution since it represents an admixture of 
formation environments which likely does not match the composition of the field. 
As we show in Figure 6, the binary frequency appears to be fundamentally 
different in the dark cloud complexes (Taurus and Chamaeleon) than in USco-A. 
This distinction suggests that binary formation processes can vary significantly 
between different environments, and therefore that analysis of other binary 
properties should take the environment into account when possible.

\begin{figure*} 
\epsscale{0.90} 
\plotone{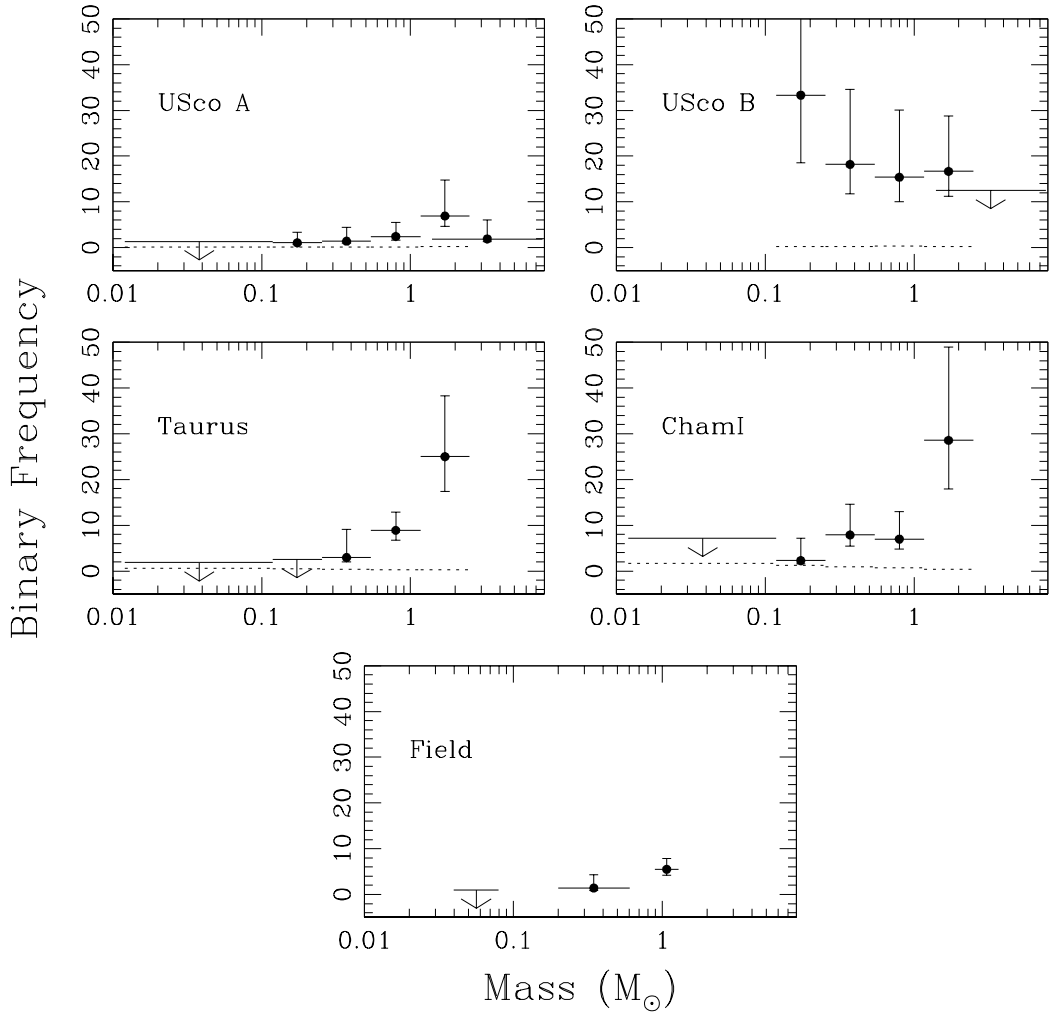} 
\caption{
The wide (330-1650 AU) binary frequency as a function of mass for each region and 
as determined from field multiplicity surveys. The higher-mass histogram bins are 
equally sized in $log M$, but the three lowest-mass bins have been combined to 
illustrate the absence of any companions. The error bars are calculated assuming 
binomial statistics. The highest-mass datapoints for USco-A and USco-B denote the 
results of Kouwenhoven et al. (2005). The dashed lines show the expected 
frequency for each bin solely from foreground and background sources and 
unbound association members; they are not distinguishable from zero in most 
bins. Most upper limits for the lowest-mass bins are also very close to zero.
} 
\end{figure*}

\begin{figure} 
\epsscale{1.00} 
\plotone{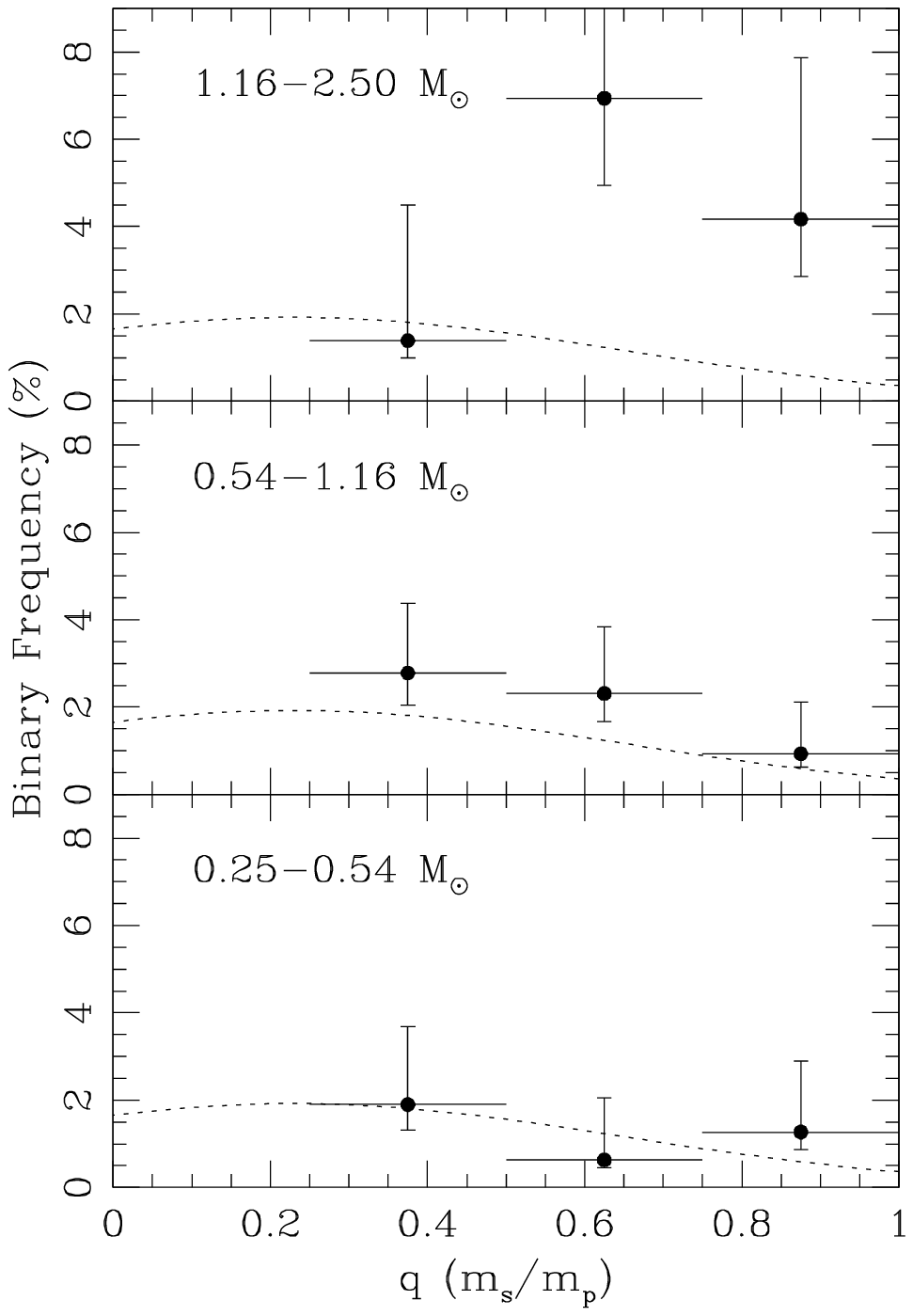} 
\caption{
The mass ratio distribution for wide binaries in the three highest-mass 
bins of our survey, calculated as a frequency among all sample members. 
The mass ratio distribution function found by Duquennoy \& Mayor (1991) 
for field solar-type stars is denoted with a dashed line. These results 
represent the sum over all associations in our sample; the binary 
frequency varies between environments (Figure 6) and our sample represents 
a different admixture of formation environments than the field sample, so 
the sample and field frequencies should be compared with caution.
} 
\end{figure}

\section{Discussion}

\subsection{The Role of Mass and Environment in Multiplicity}

Field multiplicity surveys have established several apparent trends for the mass 
dependence of binary properties. Solar-mass binaries occur at high frequency 
($\ga$60\%) and possess high mean and maximum separations (30 AU and 10$^4$ AU) 
and a mass ratio distribution biased toward low-mass companions ($q<0.5$) (e.g. 
Duquennoy \& Mayor 1991). By contrast, binaries near and below the substellar 
boundary occur at low frequency ($\sim$10-20\%) and possess small mean and 
maximum separations (4 AU and 20 AU), and a mass ratio distribution biased toward 
unity ($q$$\ga$0.7) (Close et al. 2003; Bouy et al. 2003; Burgasser et al. 2003). 
Observations of intermediate-mass M dwarfs (e.g. Fischer \& Marcy 1992; Reid \& 
Gizis 1997) suggest that their binary properties are transitional, with an 
intermediate binary frequency and possibly an intermediate separation range, plus 
a mass ratio distribution that is nearly flat for $q>0.1$.

These results have been supported by recent surveys of young open clusters and
associations (e.g. Kohler et al. 2000; Patience et al. 2002; Luhman et al. 2005;
Kraus et al. 2005, 2006; White et al. 2006). High-mass stars in these regions
typically have higher binary frequencies and wider binary separations than
lower-mass stars. There is emerging evidence that high-density regions might have
lower binary frequencies or preferentially smaller binary separations (e.g.
Kohler et al. 2006), but it has not yet been conclusively determined whether this
is a primordial feature or the result of early dynamical evolution.

\subsubsection{The Frequency of Wide Binary Formation}

Our results appear to be broadly consistent with the established paradigm of 
mass-dependent multiplicity. Wide (330-1650 AU) binaries are very common among 
stars of $\ga$1 $M_{\sun}$ and the frequency appears to decline smoothly with 
mass (Figure 6). We found few wide binaries with primaries less massive than 
$\sim$0.25 $M_{\sun}$. Wide binary systems also appear to be common in the 
low-density T associations (Taurus and Cham I), but comparatively rare in the 
USco-A OB association. This suggests that the trend against wide binaries in 
dense bound clusters might extend to unbound associations, and therefore may be 
the result of another initial condition besides stellar density.

The high frequency of wide binary systems in USco-B also suggests that binary
formation is not a pure function of stellar density. This population is
kinematically associated with the Sco-Cen complex and its proper motions most
closely match the Upper Centaurus-Lupus OB association, but the wide binary
frequency for solar-type stars in USco-B is more consistent with the T
Associations in our sample. As we discuss further in Appendix C, this
could also be explained if the stars in USco-B represent an evolved low-density
association analogous to the $\rho$ Oph or Lupus complexes rather than a 
subgroup of an OB association.

\subsubsection{The Separation Distribution of Binary Systems}

The wide binary systems discovered by our survey only represent the outer tail
of the separation distribution function. The measurement of its functional
form will require large high-resolution imaging surveys sensitive to the core
of the separation distribution ($\sim$10-100 AU for solar-type stars,
declining to $\sim$1-10 AU for brown dwarfs). The uncertainties in results
from the literature do not allow for strong constraints in this separation
range, but our results are consistent with some of the proposed environmental
trends. Wide binary systems appear to be significantly less common in USco-A
than in USco-B, a fact which was noted by Kohler et al. (2000). Their
high-resolution speckle interferometry survey found many binaries in USco-A
with projected separations of 20-300 AU, but most of the binaries they
discovered in USco-B had significantly higher separations. This led them to
conclude that the binary separation distribution is biased toward tighter 
systems in USco-A than in USco-B. Numerous multiplicity surveys in Taurus and 
Cham I (e.g. Ghez et al. 1993, 1997) have also found a wider separation 
distribution than in USco-A, which is also consistent with our results.

Field multiplicity surveys have shown a probable mass dependence in the
maximum binary separation. A census of previous surveys (Reid et al. 2001)
found that the maximum field binary separation can be described empirically
with an exponential function of the total system mass,
$a_{max}$$\propto$$10^{3.3M}$; an extension of this study to the substellar
regime by Burgasser et al. (2003) found a corresponding quadratic function,
$a_{max}$$\propto$$M^2$. Burgasser et al.  demonstrated, using the formalism
of Weinberg et al. (1987), that this is not due to interactions with field
stars or giant molecular clouds, but instead must be a feature of the
formation process or a result of early dynamical evolution in the formation
environment.

These empirical relations predict maximum separations of 330 AU and 1650 AU for 
total system masses of $\sim$0.4 and 0.6 $M_{\sun}$, respectively. This 
prediction is consistent with the general minimum primary mass of $\sim$0.25 
$M_{\sun}$ that we have identified among the wide binaries in our sample. The 
implication is that this limit is indeed set by the T Tauri stage, either as a 
result of the formation process or during dynamical evolution while these systems 
are still embedded in their natal gas cloud.

However, we have identified one candidate system, USco-160611.9-193532, with an 
apparent low-mass primary (0.13 $M_{\sun}$; SpT M5) and a very wide projected 
separation (10.8\arcsec; 1550 AU). The USNO-B proper motion for the secondary 
($\mu_{\alpha}$,$\mu_{\delta}$=-8,-18 mas yr$^{-1}$) suggests that it is a 
genuine USco member and not a background star. As we will report in a future 
publication, subsequent observations with Laser Guide Star Adaptive Optics on the 
Keck-II telescope also find that the primary is itself a close ($\sim$0.1\arcsec)  
equal-flux pair. If the wide visual companion is gravitationally bound, then this 
triple system ($M_{tot}\sim$0.4 $M_{\sun}$)  does not follow the empirical 
mass-maximum separation relations. There are several other candidate wide binary 
systems which could potentially violate these relations, but the probability of 
background star contamination is high enough in these cases that association 
membership should be confirmed via spectroscopy before any conclusions are drawn.

Finally, a census of several star-forming regions by Simon (1997) found that
pre-main sequence stars tend to cluster on two length scales, with two-point
correlation functions described by separate power laws. He concluded that
small-scale clustering is a result of binary formation, while clustering on
larger scales is a result of the condensation of multiple cores from single
molecular clouds. This could potentially explain the excess of wide companions
in Taurus, where the stars are younger and have not dispersed as far from their
formation point.  However, Simon found that the transition occurred at
separations of $\sim$$10^4$ AU in Taurus, and our survey truncates at $\sim$1500
AU. This suggests that unless his initial estimate was significantly higher than
the true transition point, all of our candidate companions fall within the
binary regime. 

\subsubsection{The Mass Ratio Distribution of Binary Systems}

Field multiplicity surveys have found that the mass ratio distribution
varies significantly with primary star mass. Most solar-mass primaries
possess binary companions with low mass ratios (Duquennoy \& Mayor 1991),
early M dwarf primaries possess companions with a uniform mass ratio
distribution (Fischer \& Marcy 1992), and late-M dwarf and brown dwarf
primaries possess companions with mass ratios near unity (e.g. Close et al.
2003; Siegler et al. 2005).

Our results can not support any strong statistical claims, but they are largely
consistent with this pattern. The only exception is that our results for the
highest-mass stars (1.16-2.50 $M_{\sun}$) suggest the presence of a possible
excess of wide similar-mass binaries. The excess is not consistent with
background star contamination since the primaries are all very bright and most
background stars should be significantly fainter; many of these binary
companions have been confirmed independently as association members. It is also
unlikely that we missed a significant number of companions with $0.25<q<0.50$;
the detection limits (in $\Delta$$K$ or $q$) are best for bright stars and
significantly exceed the limits of our statistically complete sample. This mass 
ratio distribution is also discrepant with that found by Kouwenhoven et
al.  (2005) for B-A stars in Sco-Cen; they found a distribution which is very 
similar to that of Duquennoy \& Mayor, with a strong deficit of equal-mass binaries 
compared to unequal-mass binaries. 

This discrepancy could represent another environmental dependence in multiple
star formation. Most of our high-mass binaries are found in the dark-cloud
complexes, while the Kouwenhoven sample was drawn from the three OB associations
of Sco-Cen. The field sample of Duquennoy \& Mayor is probably also dominated by
stars from dissolved OB associations or open clusters since those are thought to
be the dominant star formation channel in our galaxy. Since the binary frequency
and binary separations appear to be affected by environmental conditions, it is
plausible that the mass ratio distribution could also be affected.

\subsection{Summary of Implications for Wide Binary Formation}

Recent efforts to model binary formation have typically assumed that a cluster of 
5-10 protostellar embryos form from a single fragmenting cloud core (e.g. Kroupa 
1995; Sterzik \& Durisen 1998; Kroupa \& Bouvier 2003; Kroupa et al. 2003; 
Delgado-Donate et al. 2003; Hubber \& Whitworth 2005); these embryos would then 
undergo accretion and dynamical evolution to form single stars and stable 
multiple systems. This scenario would provide a convenient explanation for 
variations in binary properties between unbound associations or bound clusters 
since the stellar encounter rate would vary with stellar density.

However, the frequency of multiple stellar systems has been interpreted by 
Goodwin \& Kroupa (2005) to mean that the collapse and fragmentation of a cloud 
core produces only 2 or 3 stars, not 5-10. Larger systems would eject more single 
stars and tight binaries than are observed. Most models also predict that 
dynamical evolution would alter other stellar properties (spatial and velocity 
dispersion, disk lifetime, and accretion frequency), particularly at low masses.  
The preponderance of observational evidence shows that these properties are not 
consistent with strong dynamical evolution and do not vary significantly with 
mass (White \& Basri 2003; Luhman 2004d; White \& Hillenbrand 2004; Mohanty et 
al.  2005). However, some updated models of dynamical evolution suggest that not 
all of these features would show strong signatures (e.g. Bate \& Bonnell 
2005).

The apparently minor role of dynamical processes suggests that binary properties
might be established during the fragmentation of a cloud core, rather than in
subsequent dynamical evolution. Sterzik et al. (2003) suggested that the initial
cloud temperature could play a role in determining the frequency of wide binary
systems. They noted that the radius of a cloud core at the end of isothermal
collapse is inversely related to the initial cloud temperature. This suggests
that regions of low temperature will have larger spatial scales during
fragmentation, and therefore a wider distribution of binary separations. This
could be due either to a higher primordial cloud temperature or due to heating
from other young stars.

We have found that the wide binary frequency in USco-A is comparable to that of
open clusters like Praesepe or the Pleiades (Patience et al. 2002) or young
clusters like IC-348 and the ONC (Luhman et al. 2005; Kohler et al. 2006). The
stellar density is much lower in OB associations like USco-A, so the similarity
between their binary populations suggests that another initial condition might
play a key role in determining the binary separation distribution (and thus the
frequency of wide binary systems). Kohler et al. also found that binary
properties do not differ significantly between the core and the periphery of the
ONC, which span a significant range of stellar densities. These trends imply that
wide binaries only form in the absence of high-mass stars, or equivalently that
the presence of high-mass stars suppress wide binary formation. The argument by
Sterzik et al. could provide a natural explanation for this trend; high-mass
stars irradiate surrounding cloud cores, and the subsequent increase in
temperature would decrease the final length scale over which fragmentation would
occur.

\section{Conclusions}

We present the results of a search for wide binary systems among 783 members of
three nearby young associations: Taurus-Auriga, Chamaeleon I, and the two
subgroups of Upper Scorpius.  This program analyzed near-infrared $JHK$ imagery
from the Two-Micron All-Sky Survey to search for wide (1-30\arcsec;  
$\sim$150-4500 AU) companions to known association members, using color-magnitude
cuts to reject likely background stars. We identified a total of 131 candidate
binary companions in these associations, of which 39 have not been previously
identified in the literature. 

We find that the wide binary frequency (330-1650 AU; $q>0.25$) is a function of
both mass and environment, with significantly higher frequencies among high-mass
stars than lower-mass stars and in the T associations than in the OB
association. We discuss the implications for wide binary formation and conclude
that the environmental dependence is not a direct result of stellar density or
total association mass, but instead might depend on another parameter like the
gas temperature of the formation environment.

We also analyze the mass ratio distribution as a function
of mass and find that it largely agrees with the
distribution seen for field stars. There appears to be a moderate excess
of similar-mass ($q>0.5$) wide binaries among the highest-mass (1.16-2.50
$M_{\sun}$)  stars in our sample, but the number statistics do not support
any other strong conclusions. The binary populations in these associations
generally follow the empirical mass-maximum separation relation observed
for field binaries, but we have found one candidate low-mass system
(USco-160611.9-193532; $M_{tot}$$\sim$0.4 $M_{\sun}$) which has a projected
separation (10.8\arcsec; 1550 AU) much larger than the limit for its mass.

Finally, we find that the binary frequency in the USco-B subgroup is
significantly higher than the USco-A subgroup and is consistent with the measured
values in Taurus and Cham I. This discrepancy, the absence of high-mass stars in
USco-B, and its marginally distinct kinematics suggest that it might not be
directly associated with either USco-A or Upper Centaurus-Lupus, but instead
represent an older analogue of the $\rho$ Oph or Lupus associations.

\acknowledgements

The authors thank R. White and C. Slesnick for helpful feedback on the manuscript 
and on various ideas presented within.

This work makes use of data products from the Two Micron All-Sky Survey, which is a
joint project of the University of Massachusetts and the Infrared Processing and
Analysis Center/California Institute of Technology, funded by the National
Aeronautics and Space Administration and the National Science Foundation. This
research has also made extensive use of the SIMBAD database, operated at CDS,
Strasbourg, France, and of the USNOFS Image and Catalogue Archive operated by 
the United States naval Observatory, Flagstaff Station 
(http://www.nofs.navy.mil/data/fchpix/).

\appendix
\section{Comment on Proper Motions}

Photometric criteria are usually insufficient for identifying members of stellar
populations. As we show in Section 3, near-infrared color cuts allow us to
reject only $\sim$50-90\% of background and foreground contaminants. Any further
increase in the rejection rate of our survey would require either spectroscopic
observations (to test for lithium or signatures of low surface gravity, both
indicators of youth) or astrometric observations (to measure proper motions and
test for kinematic association). Performing spectroscopic observations can be a
resource-intensive undertaking, but astrometric data are now commonly available
from all-sky surveys.

The largest astrometric database currently available is the USNO-B catalogue
(Monet et al. 2003), which computed proper motions from the Palomar Observatory
Sky Surveys of the mid-1950s and early 1980s. These observations were originally
performed using wide-field photographic plates, and the USNO team digitally
scanned these plates and computed photometry and astrometry for every source. The
faint limit for photometry is $\sim$$R=19-20$, and astrometry is available for
most sources brighter than $R=17-18$. Typical proper motion uncertainties are
$\sim$2-3 mas yr$^{-1}$ in each axis for bright stars and $\la$10-15 mas yr$^{-1}$ for faint
stars.

Since the USNO-B catalogue is the result of an automated photometric/astrometric
pipeline, individual measurements are subject to some uncertainties like
distortions in the plate scale and centroid errors due to diffraction spikes from
nearby bright stars. There are also some design issues which limit its utility.
For example, objects with no proper motion information (such as from detection in
only one epoch) are reported to have proper motions of 0 mas yr$^{-1}$, but all objects
which have motions within 1$\sigma$ of zero are also rounded to 0 mas yr$^{-1}$. Thus,
it is impossible to determine whether a measurement of 0 mas yr$^{-1}$ corresponds to a
bad measurement or a genuine detection of small proper motion.  Finally, most
stars within $<$10\arcsec\, of a brighter object do not have proper motion
measurements available, so USNO-B astrometry is only potentially useful in
studying wide companions.

In Figure 8, we present a plot of the fraction of confirmed Upper Sco 
members as a function of magnitude which possess any USNO-B proper motion 
measurement (dashed line) and a measurement which lies within 15 mas 
yr$^{-1}$ of the mean association value ($\sim$3$\sigma$ for bright 
sources; solid line). The maximum fraction of confirmed members which are 
identified as kinematic members is only $\sim$2/3, and this fraction 
declines rapidly for faint targets ($K>12$). This suggests that using 
existing proper motions to select candidate binary companions would 
introduce significant incompleteness in the resulting statistics, so we 
have chosen to omit this data from our selection criteria. However, these 
proper motion measurements are useful as a test of our selection process, 
so we list the USNO-B proper motions for each candidate companion in Table 
5. Those objects with consistent proper motions could be high-priority 
candidates for spectroscopic followup or more detailed astrometric 
followup. We also list the proper motions of each primary star in our 
sample in Table 2.

\begin{figure}
\epsscale{1.00}
\plotone{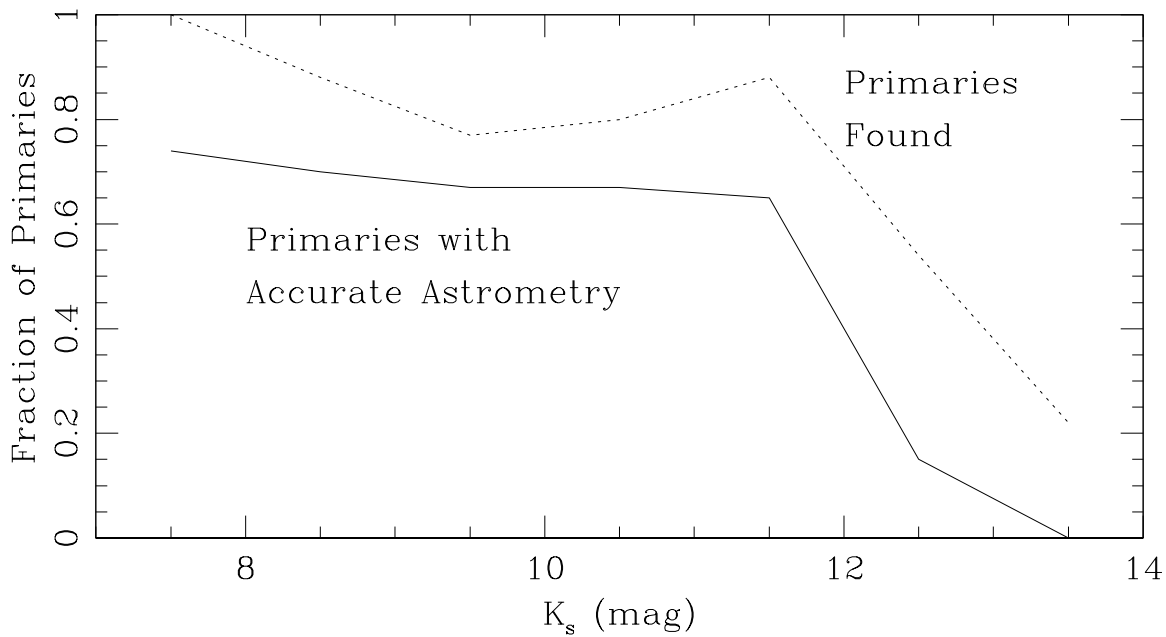}
\caption{A plot of the fraction of confirmed Upper Sco members as a function of 
magnitude which possess proper motion measurements in USNO-B (dashed line) and 
measurements which lie within 15 mas yr$^{-1}$ of the mean association value (solid 
line). The maximum fraction of members which could be recovered by kinematic 
selection criteria is only $\sim$2/3, and this declines rapidly for faint targets 
($K>12$).
} 
\end{figure}

\section{The Kinematics of Northern Sco-Cen}

The young stars of the Sco-Cen complex are divided into three subgroups: Upper
Scorpius, Upper Centaurus-Lupus, and Lower-Centaurus Crux. These three subgroups are
spatially distinct on the sky, but there is some overlap along the border between
populations. This can lead to ambiguities in assigning stars to their appropriate
population. For example, the stars of USco-B lie on the border between Upper Scorpius
and Upper-Centaurus Lupus. It is not known which group they are associated with, or
if they form another distinct population. The color-magnitude sequences for each
subgroup are not distinct due to their similar age and distance, so photometry does
not provide a reliable diagnostic of subgroup membership. However, studies of the
high-mass stars of Sco-Cen (e.g. de Zeeuw et al. 1999) have found that the space
velocities of each subgroup differ by a few km s$^{-1}$. This difference is not measurable
in the proper motions of individual stars, but it might be detected as a difference
in the mean proper motion for a population.

In Figure 9, we present proper motion diagrams for USco members which have
previously been assigned to USco-A or USco-B by Brandner et al. (1996) and
Kohler et al. (2000). The mean proper motions for each subgroup are not
directly comparable due to projection effects, but given the small radial
velocity of USco-A (-4.6 km s$^{-1}$; de Zeeuw et al. 1999) and the
locations of the association centers (16h, -22$^o$ for USco-A; 15.5h,
-31$^o$ for USco-B), the difference in proper motions should be no more than
2-3 mas yr$^{-1}$ and the vectors should be almost parallel.  We find that the
proper motion of USco-B (33.2 mas yr$^{-1}$) is significantly higher than that of
USco-A (22.3 mas yr$^{-1}$), and the vectors diverge by $\sim$15 degrees.

This result suggests not only that it is appropriate to consider USco-A
and USco-B separately for statistical purposes, but that it might be
prudent to question the relationship between USco-B and the rest of
Sco-Cen. The difference in space velocities between USco-B and the other
nearby Sco-Cen subgroups ($\sim$10 km s$^{-1}$) is far higher than that between
the major subgroups. However, any further investigation is beyond the
scope of this work.

We also conclude that the kinematic information lacks sufficient precision 
to distinguish the subgroup membership of individual stars and identify 
the boundary between the regions. Indeed, it is likely that there is no 
precise dividing line. The spatial distribution of these objects on the 
celestial sphere is only a projection of their three-dimensional 
distribution, so it is quite likely that the projected two-dimensional 
distributions overlap. This suggests that any difference between these two 
populations could be averaged out by cross-contamination. However, the 
distinct proper motions apparent in Figure 9 imply that most of the stars 
have been classified in the appropriate group.

\begin{figure} 
\epsscale{0.90} 
\includegraphics{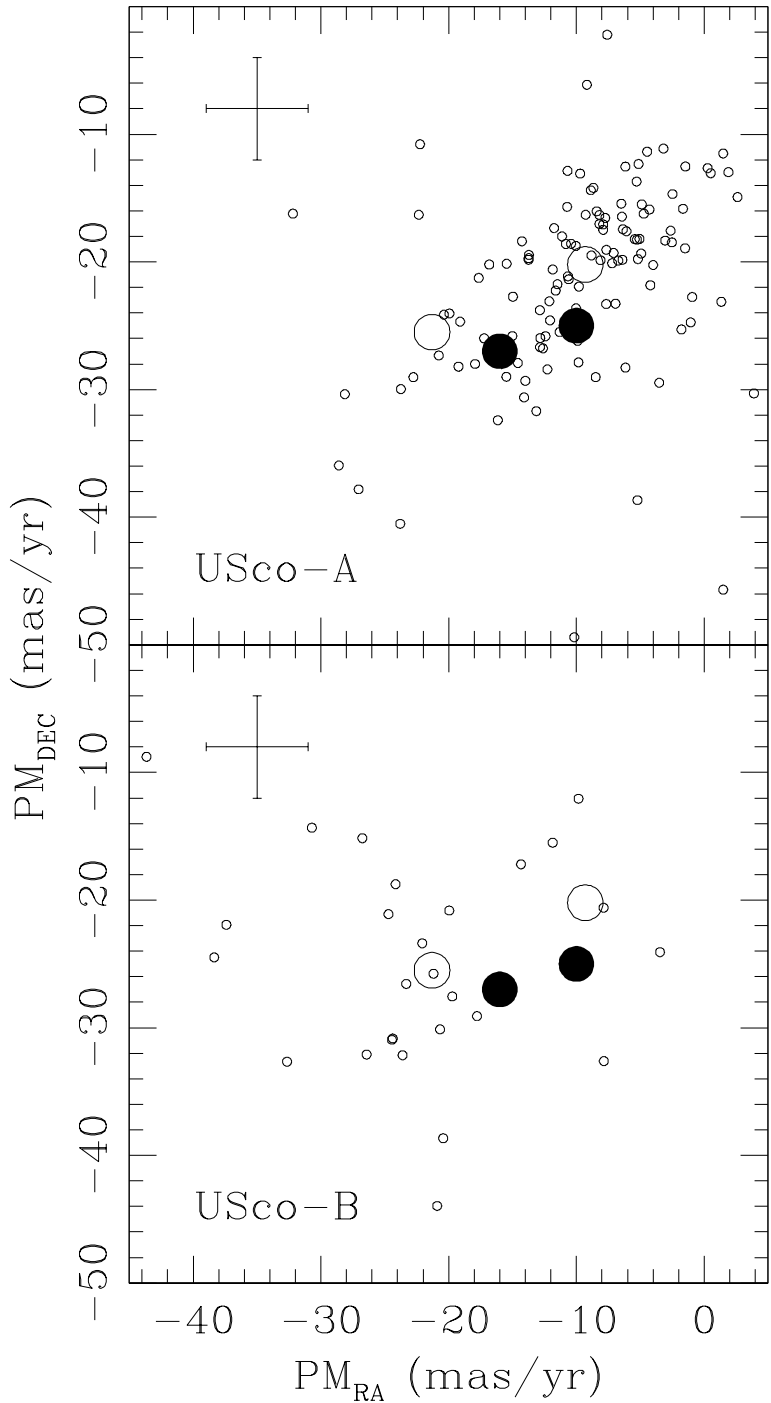} 
\caption{Proper-motion diagrams for Sco-Cen members brighter than $K=10$ which 
have been previously assigned to either USco-A or USco-B. The large filled 
circles denote the regional proper motions for Upper Sco (-10,-25) and UCL 
(-16,-27) as determined by HIPPARCOS for early-type members (de Zeeuw et al. 
1999). The large open circles denote regional proper motions for USco-A 
(-9.3,-20.2) and USco-B (-21.3,-25.5) as determined from our data. The typical 
uncertainties for individual measurements are shown with error bars in the upper 
left corner; the scatter for USco-A appears to be consistent with these 
uncertainties, but the scatter for USco-B is significantly larger. The 
uncertainties in the mean values are $\sim$0.5 mas yr$^{-1}$ for USco-A and $\sim$1.5 
mas yr$^{-1}$ for USco-B.
} 
\end{figure}

\section{The Nature of Upper Scorpius B}

The distinct binary properties observed for USco-A and USco-B suggest that
it might be prudent to reconsider the nature of USco-B. The Sco-Cen complex
consists primarily of three kinematically associated OB associations:  
Lower Centaurus-Crux, Upper Centaurus-Lupus, and Upper Scorpius. The $\rho$
Oph dark cloud complex is also associated with Sco-Cen (specifically with
USco), and the Lupus dark clouds could be kinematically associated, but the
evidence of is not yet conclusive. LCC and UCL appear to be $\sim$5-10 Myr
older than USco, which in turn is $\sim$5 Myr older than $\rho$ Oph. This
has been cited as evidence (e.g.  Mamajek et al. 2002; Sartori et al. 2003)
that triggered star formation is occurring in Sco-Cen. The implication is
that supernovae originating from the highest-mass members of UCL triggered
star formation in USco, and in turn one or more supernovae in USco triggered
star formation in $\rho$ Oph. Since the ages of UCL and LCC are somewhat
uncertain, it is unclear whether they are coeval or one triggered star
formation in the other.

USco-B is located on the border between UCL and USco, in a region largely bereft
of high-mass stars. Its age and distance are difficult to assess since there are
no high-mass members which might possess HIPPARCOS distances, but the
association's color-magnitude sequences lie slightly lower than USco-A and are
consistent with the slightly larger distance and older age of UCL. However, its
kinematics (Appendix A) appear to be marginally inconsistent with the OB
subgroups of Sco-Cen, with a spatial velocity which differs by $\sim$10 km s$^{-1}$.
Finally, its binary properties are inconsistent with the one OB association
which has been extensively studied (USco-A); no comparison is possible with UCL
since there have been no large-scale surveys for new low-mass members.

The absence of high-mass stars and high wide binary frequency in USco-B are much
more consistent with low-density T associations. This suggests that perhaps
USco-B is an older analogue to the $\rho$ Oph or Lupus clouds: an association
consisting primarily of low-mass stars whose formation was triggered by
supernovae in UCL, much as the current star formation in $\rho$ Oph was
triggered by supernovae in USco, but which is not directly associated.  
Unfortunately, it will be difficult to test this assertion. Any primordial gas
in USco-B has been dispersed, either consumed in star formation or swept away by
supernovae and stellar winds, so it only consists of an unbound association of
pre-main-sequence stars. The low galactic latitude of USco-B also results in
significant contamination from background stars, which will confuse any
photometric surveys that attempt to identify these stellar members of the
association.

\clearpage

\LongTables

\begin{deluxetable*}{lccccccccccc} 
\tabletypesize{\scriptsize}
\tablewidth{0pt} 
\tablecaption{Close Pairs of Confirmed Association Members\label{tbl3_01}} 
\tablehead{\colhead{Name} & \multicolumn{3}{c}{Primary} & 
\multicolumn{3}{c}{Secondary} & \colhead{Projected} & 
\colhead{Position}
\\
\colhead{} & \colhead{$J-K$} & \colhead{$H-K$} & \colhead{$K$} & 
\colhead{$J-K$} & \colhead{$H-K$} & \colhead{$K$} &
\colhead{Sep(\arcsec)} & \colhead{Angle(deg)} 
} 
\startdata 
2M11103-7722&2.00&0.68&10.03&2.06&0.70&10.67&9.51&145.7\\
ISO350\\
ChaHa4&1.14&0.41&11.02&1.06&0.33&13.24&20.83&128.5\\
ChaHa10\\
ChaHa10&1.06&0.33&13.24&1.04&0.37&13.55&19.60&58.9\\
ChaHa11\\
CHX18N&1.34&0.52&7.77&1.60&0.64&8.87&24.38&255.7\\
T49\\
CHXR14 N&0.94&0.23&9.60&0.98&0.23&9.75&28.17&166.6\\
CHXR14 S\\
CHXR20&1.30&0.32&8.88&1.42&0.40&9.39&28.46&349.2\\
T22\\
CHXR30 A&2.71&0.83&9.09&3.92&1.54&9.95&9.93&295.4\\
CHXR30 B\\
CHXR60&0.99&0.28&10.58&1.06&0.31&10.80&28.28&220.6\\
Hn18\\
CHXR68 A&0.92&0.24&8.87&0.98&0.27&10.26&4.39&212.4\\
CHXR68 B\\
ESO-Ha-566&1.32&0.42&11.03&1.91&0.74&14.14&23.65&93.9\\
CHSM15991\\
Hn10E&1.91&0.69&10.05&3.80&1.42&10.00&19.17&231.0\\
C1-25\\
Hn21 W&1.34&0.44&10.65&1.27&0.48&11.49&5.43&69.3\\
Hn21 E\\
ISO143&1.48&0.56&11.10&1.02&0.43&13.04&18.16&223.8\\
ISO138\\
ISO237&2.31&0.82&8.62&1.33&0.40&9.24&28.32&235.7\\
T45A\\
T28&1.91&0.72&8.26&1.27&0.49&11.51&28.87&164.3\\
ChaHa8\\
T29&2.67&1.09&6.83&3.33&1.37&8.30&16.37&81.8\\
ESO-Ha-562\\
T31&1.74&0.68&6.96&2.38&0.97&9.89&16.52&221\\
T30\\
T34&1.17&0.32&10.02&1.12&0.4&10.67&25.41&3.9\\
ChaHa13\\
T47&1.97&0.78&9.18&1.30&0.35&10.75&12.09&161.3\\
ESO-Ha-568\\
T52&1.44&0.62&6.85&1.79&0.73&9.13&11.18&99.2\\
T53\\
DHTau&1.59&0.65&8.18&0.93&0.21&8.39&15.23&126\\
DITau\\
FVTau&2.48&0.88&7.44&1.93&0.62&8.87&12.29&105.7\\
FVTau/c\\
FZTau&2.55&1.05&7.35&1.93&0.62&8.05&17.17&250.5\\
FYTau\\
GGTau A&1.31&0.45&7.36&1.09&0.42&9.97&10.38&185.1\\
GGTau B\\
GHTau&1.32&0.44&7.79&1.19&0.40&6.96&21.77&15.2\\
V807Tau\\
GKTau&1.59&0.64&7.47&1.45&0.53&7.89&13.14&328.4\\
GITau\\
HBC352&0.51&0.14&9.58&0.59&0.14&9.86&8.97&70.8\\
HBC353\\
HBC355&0.62&0.13&10.20&0.73&0.15&11.11&6.31&298.3\\
HBC354\\
HLTau&3.21&1.76&7.41&2.09&0.86&7.29&23.31&91.2\\
XZTau\\
HPTau-G2&0.87&0.26&7.23&1.92&0.84&7.63&21.30&296.9\\
HPTau\\
HPTau-G2&0.87&0.26&7.23&1.24&0.36&8.80&10.09&243.4\\
HPTau-G3\\
2MASSJ04554757+3028077&1.07&0.33&9.98&1.03&0.43&12.16&6.31&115.7\\
2MASSJ04554801+3028050\\
LkHa332-G1&1.64&0.46&7.95&1.56&0.44&8.23&25.88&254.5\\
LkHa332-G2\\
LkHa332-G2&1.64&0.46&7.95&1.87&0.66&7.94&10.51&35.3\\
V955Tau\\
MHO-2&3.73&1.63&7.79&4.70&2.10&7.78&3.93&333.9\\
MHO-1\\
V773TauA&1.28&0.43&6.21&1.52&0.69&11.64&23.38&215.9\\
2MASSJ04141188+2811535\\
V928Tau&1.43&0.33&8.11&1.16&0.41&10.38&18.25&228.2\\
CFHT-Tau-7\\
RXJ1524.2-3030A&0.63&0.16&8.68&0.99&0.29&9.60&20.18&87.3\\
RXJ1524.2-3030B\\
RXJ1537.0-3136A&0.52&0.04&7.74&0.90&0.18&7.65&5.37&285.0\\
RXJ1537.0-3136B\\
RXJ1539.4-3446B&1.21&0.43&7.98&1.61&0.60&9.29&6.36&98.1\\
RXJ1539.4-3446C\\
RXJ1540.7-3121A&0.83&0.26&10.53&0.86&0.27&10.66&5.95&75.5\\
RXJ1540.7-3121B\\
RXJ1558.1-2405A&0.79&0.16&8.96&0.89&0.25&11.06&18.15&254.4\\
RXJ1558.1-2405B\\
RXJ1604.3-2130A&1.44&0.60&8.51&1.02&0.27&9.43&16.22&215.9\\
RXJ1604.3-2130B\\
USco-160428.4-190441&1.04&0.27&9.28&1.02&0.31&11.01&9.77&321.3\\
USco-160428.0-190434\\
USco-160707.7-192715&0.95&0.24&9.80&1.00&0.29&11.17&23.45&140.4\\
USco-160708.7-192733\\
USco-160822.4-193004&0.97&0.18&9.06&1.11&0.27&9.47&13.47&71.4\\
USco-160823.2-193001\\
USco-160900.7-190852&1.07&0.32&9.15&1.01&0.38&10.96&18.92&326.5\\
USco-160900.0-190836\\
USco-161010.4-194539&0.97&0.28&10.41&0.96&0.33&11.38&25.59&160.8\\
USco-161011.0-194603\\
ScoPMS008b&0.97&0.32&9.33&1.03&0.38&9.77&25.61&68.6\\
ScoPMS008a\\
\enddata 
\end{deluxetable*}

\clearpage
\begin{landscape}
\begin{deluxetable*}{lccccccccccl} 
\tabletypesize{\tiny}
\tablewidth{0pt} 
\tablecaption{Candidate Wide Binary Systems(Full Table)\label{tbl3}} 
\tablehead{\colhead{Name} & \multicolumn{3}{c}{Primary} & 
\multicolumn{3}{c}{Secondary} & \colhead{Projected} & 
\colhead{Position} & \colhead{$\mu_{\alpha}$,$\mu_{\delta}$\tablenotemark{a}} &
\colhead{Ident} & \colhead{References}
\\
\colhead{} & \colhead{$J-K$} & \colhead{$H-K$} & \colhead{$K$} & 
\colhead{$J-K$} & \colhead{$H-K$} & \colhead{$K$} &
\colhead{Sep(\arcsec)} & \colhead{Angle(deg)} & \colhead{(mas yr$^{-1}$)} & 
\colhead{Method}
} 
\startdata 
2M11103-7722&2.00&0.68&10.03&2.21&0.77&13.85&9.30&108.8&-&PSC&-\\
C7-1&1.78&0.62&10.55&1.67&0.43&13.32&5.73&214.9&0,0&PSC&-\\
CHSM1715&2.05&0.85&10.9&1.42&0.43&13.94&9.07&30.3&-58,42&PSC&-\\
CHXR26&2.02&0.46&9.92&2.68&1.07&9.98&1.41&215.2&-&PSF&Luhman (2004b)\\
CHXR28&1.17&0.32&8.23&1.53&0.39&8.83&1.78&121.6&-&PSF&Brandner et al. (1996)\\
CHXR9C&1.01&0.24&8.95&1.15&0.54&13.46&4.53&333.0&-&PSF&-\\
KG102&1.15&0.34&12.01&1.29&0.58&13.05&2.24&223.7&-&PSF&Persi et al. 2005\\
T3&2.68&1.15&8.87&1.29&0.41&10.35&2.22&290.7&-&PSF&Reipurth \& Zinnecker (1993)\\
T6&1.50&0.65&7.76&1.38&0.50&11.15&4.99&123.0&0,0&PSC&Ghez et al. (1997)\\
T14A&1.88&0.79&12.45&2.38&1.26&13.56&2.5&52.5&-&PSF&Haisch et al. (2004)\\
T26&1.60&0.73&6.22&1.13&0.43&7.28&4.16&203.4&-24,6&PSC&Reipurth \& Zinnecker (1993)\\
T33A+B&1.91&0.89&7.22&1.89&1.09&7.76&2.46&286.0&-&PSF&Chelli et al. (1988)\\
T39&0.97&0.19&8.96&1.03&0.29&9.98&4.17&77.1&0,0&PSC&Reipurth \& Zinnecker (1993)\\
T51&1.05&0.42&8.27&1.75&0.63&10.04&1.98&161.9&-&PSF&Reipurth \& Zinnecker (1993)\\
2MASSJ04080782+2807280&1.06&0.35&11.39&0.77&0.21&9.34&9.43&351.1&0,0&PSC&-\\
2MASSJ04414489+2301513&1.26&0.57&13.16&0.88&0.24&9.85&12.37&57.3&-2,-18&PSC&-\\
CIDA-9&1.68&0.6&11.49&1.38&0.56&12.15&2.33&59.3&-&PSF&White et al. (2006)\\
CoKuTau3&2.22&0.66&8.66&2.48&1.10&9.91&2.07&174.9&-&PSF&Leinert et al. (1993)\\
CoKuTau3&2.32&0.79&8.41&2.77&0.81&13.38&12.60&349.2&-&PSC&-\\
DKTau&1.27&0.52&7.78&1.75&0.57&8.38&2.37&119.7&-&PSF&Leinert et al. (1993)\\
FWTau&0.95&0.29&9.39&1.37&0.38&9.42&12.22&246.7&0,0&PSC&Hartmann et al. (2005)\\
GGTauB&0.97&0.37&10.29&1.53&0.65&11.39&1.55&130.3&-&PSF&Leinert et al. (1993)\\
Haro6-37&1.70&0.69&7.76&1.77&0.60&8.58&2.70&37.9&-&PSF&Leinert et al. (1993)\\
HBC356&0.57&-0.08&10.82&0.83&0.49&10.9&1.17&1.1&-&PSF&White et al. (2006)\\
HBC427&0.83&0.19&8.13&0.77&0.19&9.02&14.9&154.0&0,0&PSC&-\\
HNTau&2.36&1.08&8.40&1.13&0.68&11.59&3.10&218.7&-&PSF&Leinert et al. (1993)\\
HVTau&1.26&0.48&7.94&1.92&0.89&12.29&3.76&43.9&-&PSF&Simon et al. (1995)\\
ISTau&1.68&0.65&8.64&1.87&0.53&13.46&10.85&57.4&18,-302&PSC&-\\
ITTau&1.94&0.81&8.07&1.62&0.34&9.81&2.37&223.4&-&PSF&White \& Ghez (2001)\\
J1-4872&1.14&0.30&8.56&1.21&0.48&9.25&3.38&232.9&-&PSF&Reipurth \& Zinnecker (1993)\\
JH112&2.07&0.83&8.17&1.91&0.61&9.20&6.56&34.3&0,0&PSC&White et al. (2006)\\
JH223&1.22&0.43&9.52&1.17&0.4&12.19&2.06&342.3&-&PSF&White et al. (2006)\\
LkCa4&0.93&0.20&8.32&1.53&0.42&13.57&8.86&154.6&-310,-134&PSC&-\\
LkCa7&0.89&0.77&8.24&0.54&-2.58&12.04&1.18&25.1&-&PSF&Leinert et al. (1993)\\
UZTau&1.78&0.76&7.35&1.94&0.53&7.47&2.80&275.8&-10,-20&PSC&White \& Ghez (2001)\\
V410-Xray5a&1.84&0.63&10.15&2.12&0.71&13.82&13.27&47.7&-&PSC&-\\
V710Tau&0.63&0.45&8.65&0.46&0.29&8.52&3.03&178.5&8,-28&PSC&Leinert et al. (1993)\\
GSC 06785-00476&0.59&0.14&8.92&0.87&0.28&11.96&6.30&82.6&-&PSC&-\\
GSC 06204-01067&0.97&0.22&8.75&1.09&0.39&10.56&2.49&89.2&-&PSF&-\\
GSC 06780-01061&0.89&0.39&9.06&1.04&-0.13&10.36&1.50&270.3&-&PSF&-\\
GSC 06784-00039&0.64&0.14&7.91&1.11&0.42&13.03&13.53&77.5&-&PSC&-\\
GSC 06784-00997&0.89&0.17&8.36&0.91&0.31&11.26&4.81&240.4&-16,-32&PSC&-\\
GSC 06213-00306&0.93&0.24&8.59&1.04&0.33&10.73&3.18&305.5&-&PSF&-\\
GSC 06793-00868&0.88&0.19&9.25&1.06&0.22&9.59&2.01&156.5&-&PSF&-\\
GSC 06793-00806&1.31&0.60&8.26&1.10&0.15&9.31&1.89&342.4&-&PSF&Gregorio-Hetem 1992\\
RXJ1555.8-2512&0.46&0.10&8.29&0.93&0.34&12.53&14.61&298.1&74,316&PSC&-\\
RXJ1555.8-2512&0.46&0.10&8.29&0.96&0.27&10.00&8.91&318.4&0,0&PSC&-\\
RXJ1558.8-2512&0.90&0.21&9.65&0.92&0.26&11.53&11.35&130.1&0,0&PSC&-\\
RXJ1559.2-2606&0.72&0.11&9.41&0.94&0.31&10.65&2.96&328.3&-&PSF&Kohler et al. (1999)\\
RXJ1600.7-2343&0.95&0.02&10.81&0.92&0.54&10.89&1.41&28.3&-&PSF&Kohler et al. (1999)\\
RXJ1602.8-2401B&0.80&0.16&8.93&0.92&0.34&11.62&7.22&352.9&-&PSC&-\\
RXJ1606.6-2108&1.30&0.63&9.43&0.61&-0.19&10.27&1.17&28.2&-&PSF&Kohler et al. (1999)\\
SCH16151115-24201556&1.06&0.42&13.17&0.96&0.30&12.13&17.96&69.8&0,0&PSC&-\\
ScoPMS008a&0.93&0.32&10.14&0.98&0.14&10.97&1.58&95.4&-&PSF&Kohler et al. (1999)\\
ScoPMS016&1.04&0.19&9.59&0.86&0.30&10.00&1.37&45.0&-&PSF&Kohler et al. (1999)\\
ScoPMS042b&1.06&0.28&9.62&1.05&0.42&11.93&4.58&6.8&-8,-6&PSC&Kohler et al. (1999)\\
ScoPMS048&0.89&0.23&8.09&0.80&0.13&8.34&3.05&192.1&0,0&PSC&Kohler et al. (1999)\\
ScoPMS052&0.82&0.17&7.49&1.17&0.36&9.11&19.06&269.5&4,-18&PSC&Martin et al.(1998)\tablenotemark{b}\\
USco80&0.93&0.32&12.08&0.92&0.25&10.19&12.27&15.2&0,0&PSC&-\\
USco-155532.4-230817&0.65&-0.08&12.94&1.43&0.73&13.19&1.79&207.7&-&PSF&\\
USco-160202.9-223613&0.86&0.13&11.90&1.22&0.50&12.80&2.30&94.7&-&PSF&Bouy et al. (2006)\\
USco-160258.5-225649&0.54&-0.13&10.61&2.49&1.20&11.35&1.21&59.1&-&PSF&-\\
USco-160611.9-193532&0.99&0.33&11.02&1.22&0.54&11.78&10.78&226.5&-8,-18&PSC&-\\
USco-160700.1-203309&1.17&0.34&9.94&1.03&0.30&9.54&11.65&293.1&-2,-22&PSC&-\\
USco-160702.1-201938&1.08&0.34&11.86&1.22&0.50&12.11&1.63&242.3&-&PSF&-\\
USco-160904.0-193359&0.86&0.34&11.37&1.09&0.25&11.74&1.28&328.5&-&PSF&-\\
USco-160908.4-200928&1.00&0.27&9.98&0.96&0.24&10.30&1.93&139.4&-&PSF&-\\
USco-160936.5-184800&1.26&0.48&10.28&1.24&0.32&12.50&19.97&2.2&24,-2&PSC&-\\
USco-161031.9-191305&1.03&0.26&8.99&1.22&0.45&12.73&5.71&114.0&-&PSC&-\\
USco-161039.5-191652&1.04&0.26&10.27&1.05&0.38&12.25&14.95&183.2&-66,-124&PSC&-\\
GSC 06770-00655&0.58&0.11&9.75&0.83&0.22&10.56&9.01&327.80&0,0&PSC&-\\
GSC 06770-00655&0.58&0.11&9.75&0.88&0.25&10.01&29.62&325.3&16,-10&PSC&-\\
RXJ1528.7-3117&0.79&0.31&7.5&0.39&0.31&8.06&2.46&181.8&-&PSF&Kohler et al. (1999)\\
RXJ1529.4-2850&0.73&0.66&7.71&0.68&0.26&7.87&2.07&168.3&-&PSF&Kohler et al. (1999)\\
RXJ1530.4-3218&0.51&0.36&7.74&0.78&0.39&7.83&2.07&23.0&-&PSF&Kohler et al. (1999)\\
RXJ1536.5-3246&0.86&0.22&10.26&0.87&0.25&10.54&2.37&134.9&-&PSF&Kohler et al. (1999)\\
RXJ1539.4-3446B&1.21&0.43&7.98&2.31&0.77&10.24&27.80&9.9&0,0&PSC&-\\
RXJ1539.4-3446B&1.21&0.43&7.98&3.06&0.97&13.13&29.37&79.6&-&PSC&-\\
RXJ1543.8-3306&0.91&0.34&10.24&0.82&0.14&10.64&2.79&185.1&-&PSF&Kohler et al. (1999)\\
RXJ1545.2-3417&1.11&0.54&7.04&1.23&0.81&8.36&2.60&297.2&-&PSF&Kohler et al. (1999)\\
RXJ1554.0-2920&0.96&0.14&8.87&0.43&0.25&10.81&1.44&73.8&-&PSF&Kohler et al. (1999)\\
RXJ1554.0-2920&0.91&0.25&8.74&0.74&0.19&10.61&26.33&257.5&0,0&PSC&-\\
\enddata
\tablenotetext{a}{An entry of 0,0 denotes a source which was detected by 
the USNO-B survey, but did not show a significant proper motion. An entry of "-" 
denotes a source which was not detected by the USNO-B survey.}
\tablenotetext{b}{ScoPMS052 B is also known as GSC06209-01312; Martin et al. (1998) 
identified it as a WTTS.}
\end{deluxetable*}

\clearpage

\begin{deluxetable*}{lcccccccccl} 
\tabletypesize{\tiny}
\tablewidth{0pt}
\tablecaption{Ultrawide Visual Companions(Full Table)\label{tbl3}} 
\tablehead{\colhead{Name} & \multicolumn{3}{c}{Primary} & 
\multicolumn{3}{c}{Secondary} & \colhead{Projected} & 
\colhead{Position} & \colhead{$\mu_{\alpha}$,$\mu_{\delta}$} &
\colhead{References}
\\
\colhead{} & \colhead{$J-K$} & \colhead{$H-K$} & \colhead{$K$} & 
\colhead{$J-K$} & \colhead{$H-K$} & \colhead{$K$} &
\colhead{Sep(\arcsec)} & \colhead{Angle(deg)} & \colhead{(mas yr$^{-1}$)}
} 
\startdata 
C1-6&3.92&1.68&8.67&1.93&0.80&14.10&27.58&156.0&-&OTS12(candidate; Oasa et al. 1999)\\
C1-6&3.92&1.68&8.67&2.26&0.70&13.75&24.51&123.8&-&OTS14(candidate; Oasa et al. 1999)\\
Cam2-19&2.40&0.74&10.25&2.36&0.72&13.45&23.13&107.6&-&-\\
Cam2-42&2.44&0.73&9.16&2.09&0.50&13.51&27.64&261.7&-&-\\
Cam2-42&2.44&0.73&9.16&2.11&0.65&14.14&28.18&180.4&-&-\\
ChaHa11&1.04&0.37&13.55&1.35&0.31&12.50&20.96&22.1&20,-14&-\\
ChaHa7&1.19&0.48&12.42&2.65&0.85&9.88&14.47&347.4&0,0&[CCE98] 2-26(candidate; Cambresy et al. 1998)\tablenotemark{a}\\
CHSM10862&1.61&0.67&12.33&0.94&0.15&10.39&14.22&184.6&0,0&-\\
CHX18N&1.34&0.52&7.77&0.99&0.24&7.46&29.85&154.1&0,0&-\\
CHXR15&1.02&0.38&10.24&2.73&0.99&13.96&26.27&164.1&-&-\\
CHXR22E&1.89&0.57&10.00&1.58&0.48&12.48&10.65&301.7&0,0&CHXR22W(background; Luhman 2004b)\\
CHXR26&2.25&0.70&9.35&2.78&0.87&10.86&24.20&3.9&0,0&[CCE98] 2-27(candidate; Cambresy et al. 1998)\\
CHXR28&1.52&0.36&7.69&1.59&0.45&13.6&19.82&78.7&0,0&-\\
CHXR30A&2.71&0.83&9.09&1.85&0.57&12.01&25.72&56.5&2,12&-\\
CHXR35&0.98&0.35&10.87&1.28&0.51&13.4&11.23&321.5&0,0&-\\
CHXR40&1.11&0.27&8.96&0.58&0.18&7.85&28.99&129.3&-26,10&CHX15A(candidate; Luhman 2004b)\\
CHXR47&1.46&0.41&8.28&1.34&0.37&13.35&13.62&251.9&0,0&-\\
CHXR54&0.91&0.22&9.5&1.81&0.69&13.73&27.26&316.7&36,-12&-\\
CHXR74&1.23&0.30&10.21&1.46&0.59&14.12&11.27&128.2&0,0&-\\
CHXR76&1.17&0.32&10.95&2.65&0.85&9.88&27.86&169.4&0,0&[CCE98] 2-26(candidate; Cambresy et al. 1998)\tablenotemark{a}\\
CHXR78C&1.09&0.33&11.22&1.26&0.33&8.29&20.86&19.1&0,0&CHXR78NE(background; Luhman 2004b)\\
CHXR79&2.59&1.05&9.07&1.96&0.59&13.05&17.72&289.8&0,0&-\\
ESO-Ha-560&1.21&0.37&11.03&1.89&0.61&13.32&23.06&182&0,0&-\\
ESO-Ha-569&1.38&0.48&14.58&2.19&0.85&14.16&23.93&170.9&-&-\\
Hn11&2.33&0.82&9.44&2.18&0.76&14.04&25.52&167.5&-&OTS36(candidate; Oasa et al. 1999)\\
Hn11&2.33&0.82&9.44&2.88&1.11&13.77&18.08&356.4&-&OTS32(candidate; Luhman 2004b)\\
Hn12W&0.95&0.33&10.78&1.64&0.58&13.48&29.78&296.6&0,0&-\\
Hn5&1.44&0.60&10.13&0.78&0.24&9.35&23.68&172.1&0,0&-\\
Hn5&1.44&0.60&10.13&1.39&0.39&13.14&17.74&51.3&-26,10&-\\
ISO165&1.62&0.62&11.44&1.33&0.57&12.96&14.46&230.6&-&ChaI737(candidate; Lopez-Marti et al. 2004)\\
ISO237&2.31&0.82&8.62&2.00&0.71&13.36&23.96&342.5&-&OTS42(background;Luhman 2004b)\\
ISO256&2.93&1.17&11.34&2.03&0.74&14.06&28.69&244.8&-&-\\
KG102&1.26&0.43&11.80&1.16&0.40&13.49&18.82&312.3&0,0&KG102-Anon1(candidate; Persi et al. 2005)\\
OTS44&1.75&0.77&14.67&2.33&0.70&13.27&27.75&17&0,0&OTS46(candidate; Oasa et al. 1999)\\
T11&0.91&0.25&8.20&2.36&0.74&12.42&21.65&152.3&0,0&-\\
T11&0.91&0.25&8.20&2.51&0.73&14.21&28.31&57&-&-\\
T21&1.18&0.41&6.42&1.94&0.67&13.6&21.90&246.6&-&NIR9(candidate; Persi et al. 2001)\\
T27&1.14&0.39&9.52&1.02&0.30&12.61&24.70&315.4&0,0&-\\
T42&3.01&1.34&6.46&3.47&1.20&11.97&27.84&314.6&-&Cam2-44(background; Luhman 2004b)\\
T43&2.04&0.75&9.25&1.45&0.39&13.59&22.88&254.4&-30,40&-\\
T46&1.46&0.51&8.45&1.09&0.32&12.89&27.88&136.5&0,0&-\\
T51&1.28&0.52&8.00&0.94&0.15&8.57&11.40&65.7&-&CHX20A(background; Luhman 2004b)\\
2MASSJ04161885+2752155&1.19&0.43&11.35&1.12&0.28&11.95&28.04&218.2&-10,6&-\\
2MASSJ04213460+2701388&1.46&0.53&10.44&1.62&0.53&13.14&17.18&265.7&0,0&-\\
CFHT-Tau-4&1.84&0.68&10.33&2.41&0.71&13.89&24.40&72.9&-&-\\
CFHT-Tau-7&1.16&0.41&10.38&0.81&0.25&11.20&21.76&207.2&0,0&JH90(candidate; Jones \& Herbig 1979)\\
CFHT-Tau-21&2.57&1.03&9.01&1.44&0.45&11.07&23.31&152.1&0,0&-\\
DGTau&1.70&0.73&6.99&2.27&0.69&13.7&16.43&234.3&-&-\\
DOTau&2.17&0.94&7.3&3.01&1.05&10.58&28.75&8.4&-&-\\
FMTau&1.57&0.63&8.76&2.76&0.81&13.74&26.21&91.7&-&-\\
FOTau&1.53&0.45&8.12&1.62&0.48&14.10&26.19&250.8&-12,22&-\\
FSTau&2.53&1.07&8.18&3.33&1.60&11.75&19.88&275.8&64,22&Haro 6-5 B\tablenotemark{b}\\
GMAur&1.06&0.32&8.28&0.73&0.15&8.56&28.31&202.2&0,0&-\\
I04158+2805&2.60&1.17&11.18&1.27&0.39&12.16&25.34&28.9&0,0&-\\
I04216+2603&1.74&0.70&9.05&1.24&0.39&12.71&27.96&337.0&34,-16&-\\
I04385+2550&2.65&0.92&9.20&1.79&0.51&12.23&18.94&343.3&2,4&-\\
IPTau&1.43&0.54&8.35&1.16&0.38&13.43&15.75&55.7&228,-164&NLTT 13195(foreground; Salim \& Gould 2003)\\
ISTau&1.68&0.65&8.64&1.75&0.46&14.28&28.73&261.1&-4,12&-\\
LkCa15&1.26&0.44&8.16&0.96&0.20&7.02&27.62&4.6&6,4&-\\
MHO-2&3.73&1.63&7.79&3.28&1.11&12.11&26.32&269.9&-&-\\
V410-Anon20&4.47&1.48&11.93&4.45&1.44&12.55&22.71&115.3&-&V410-Anon21(background; Luhman 2000)\\
V410-Xray1&1.94&0.65&9.08&1.45&0.51&11.74&27.95&137.4&0,0&-\\
V410-Xray2&4.56&1.49&9.22&4.04&1.55&13.69&17.72&105.6&-&-\\
V410-Xray6&1.40&0.47&9.13&2.61&0.81&13.35&26.49&34.4&-&-\\
V710Tau&0.63&0.45&8.65&2.22&0.89&10.04&27.97&105.7&10,-20&-\\
DENIS-P-J162041.5-242549.0&1.49&0.52&12.9&1.32&0.35&11.62&26.73&164.5&4,12&-\\
SCH16075850-20394890&1.01&0.37&12.59&3.25&1.50&7.81&21.52&200.7&-4,-26&The (1964)\tablenotemark{d}\\
SCH16075850-20394890&1.01&0.37&12.59&1.90&0.92&13.98&22.94&285.5&0,0&-\\
SCH16182501-23381068&1.28&0.44&12.45&1.31&0.34&12.25&24.73&229.1&0,0&-\\
SCH16213591-23550341&1.22&0.46&12.73&1.42&0.38&12.54&25.65&165.3&0,0&-\\
USco-160245.4-193037&0.99&0.31&11.14&1.14&0.58&13.88&28.19&72.9&40,-26&-\\
USco-160428.4-190441&1.04&0.27&9.28&0.78&0.19&9.79&24.15&134.3&-2,4&Field; Preibisch et al. (1998)\tablenotemark{c}\\
\enddata 
\tablenotetext{a}{The source [CCE98] 2-26 is an ultrawide neighbor of both ChaHa7 and CHXR76; its 
physical association, if any, is uncertain.}
\tablenotetext{b}{Haro 6-5 B is a known member of Taurus (Mundt et al. 1984), but was not included as 
part of our statistical sample because its spectral type is uncertain.}
\tablenotetext{c}{USco-160428.4-190441 B is also known as GSC06208-00611; Preibisch et al. (1998) 
identified it as a field star.}
\tablenotetext{d}{SCH16075850-20394890 B is also known as T64-2; The (1964) identified it as a strong 
H$\alpha$ emitter.}
\end{deluxetable*}
\clearpage
\end{landscape}

\begin{deluxetable*}{lccccccccll} 
\tabletypesize{\scriptsize}
\tablewidth{0pt} 
\tablecaption{Inferred Binary Properties(Full Table)\label{tbl3}} 
\tablehead{\colhead{Name} & \multicolumn{2}{c}{Primary}
& \multicolumn{2}{c}{Secondary} & \colhead{Projected} 
& \colhead{Mass} 
\\
\colhead{}  & \colhead{SpT} & \colhead{Mass}
& \colhead{SpT} & \colhead{Mass} & 
\colhead{Separation(AU)} & \colhead{Ratio(q)}
} 
\startdata 
2M11103&M4&0.27&(M8.5)&(0.02)&1535&0.08\\
2M11103(/ISO250)&M4&0.27&M4.75(M5.5)&0.20(0.15)&1569&0.56\\
C7-1&M5&0.18&(M8)&(0.03)&945&0.18\\
CHSM1715&M4.25&0.25&(M7)&(0.05)&1497&0.18\\
CHXR9C&M2.25&0.47&(M9)&(0.02)&747&0.05\\
CHXR26&M3.5&0.33&(M5)&(0.19)&233&0.57\\
CHXR28&K6&0.77&(M3)&(0.4)&294&0.52\\
CHXR30 A&K8&0.68&M1.25(M5.5)&0.56(.14)&1638&0.2\\
CHXR68 A&K8&0.68&M2.25(M4.5)&0.54(.22)&724&0.32\\
Hn21 W&M4&0.27&M5.75(M5.5)&0.12(0.14)&896&0.52\\
KG102&M5.5&0.14&(M7)&(0.06)&370&0.41\\
T3&M0.5&0.6&(M1)&(0.56)&366&0.93\\
T6&K0&1.69&(M5)&(0.17)&823&0.1\\
T14A&K7&0.72&(M4.5)&(0.22)&413&0.3\\
T26&G2&2.34&(K0)&(1.66)&686&0.71\\
T33A+B&G7&2.14&(K0)&(1.63)&406&0.76\\
T39&M2.25&0.5&(M4.5)&(0.22)&688&0.43\\
T51&K3.5&0.93&(M5)&(0.16)&327&0.17\\
2MASSJ04080782+2807280&M3.75&0.3&(K1)&(1.47)&1367&4.9\\
2MASSJ04414489+2301513&M8.25&0.027&(M3.5)&(0.3)&1794&11\\
2MASSJ04554757+3028077&M4.75&0.20&M5.6(M7.5)&0.13(0.04)&915&0.22\\
CIDA-9&M0&0.64&(M2.5)&(0.47)&338&0.73\\
CoKuTau3&M1&0.57&(M5)&(0.17)&300&0.29\\
DKTau&K7&0.72&(M3.5)&(0.32)&344&0.45\\
FVTau&K5&0.82&M3(M2.5)&0.40(0.46)&1782&0.56\\
FWTau&M5.5&0.14&(M6)&(0.09)&1772&0.65\\
GGTau A(/B)&K7&0.72&M5(M6)&0.18(0.1)&1505&0.14\\
GGTau Bab&M5&0.18&(M7)&(0.05)&225&0.29\\
GKTau&K7&0.72&K6(M0)&0.77(0.62)&1905&0.86\\
Haro6-37&K6&0.77&(M2.5)&(0.42)&392&0.55\\
HBC352&G0&2.49&G5(G8)&2.26(2.04)&1301&0.82\\
HBC355&K2&1.2&K3(K7)&0.94(0.67)&915&0.56\\
HBC356&K2&1.2&(K3)&(0.96)&170&0.8\\
HBC427&K7&0.72&(M3)&(0.39)&2161&0.54\\
HNTau&K5&0.82&(M4.5)&(0.2)&450&0.24\\
HPTau-G2&G0&2.49&K7(K7)&0.72(0.75)&1463&0.3\\
HVTau&M1&0.57&(M8.5)&(0.02)&545&0.04\\
ISTau&K7&0.72&(M8.5)&(0.02)&1573&0.03\\
ITTau&K0&1.69&(K7)&(0.73)&344&0.43\\
J1-4872&K7&0.72&(M3)&(0.41)&490&0.57\\
JH112&K6&0.77&(M3)&(0.42)&951&0.55\\
JH223&M2&0.5&(M6.5)&(0.07)&299&0.13\\
LkCa4&K7&0.72&(M9)&(0.01)&1285&0.02\\
LkCa7&K7&0.72&(M7)&(0.05)&171&0.07\\
LkHa332-G2/V955 Tau&K7&0.72&K7(M0)&0.72(0.64)&1524&0.89\\
MHO-2&M2.5&0.45&M2.5(M4.5)&0.45(0.21)&570&0.47\\
UZTau&M1&0.57&(M2.5)&(0.46)&406&0.81\\
V410-Xray5a&M5.5&0.14&(M9)&(0.01)&1924&0.1\\
V710Tau&M1&0.57&(M2)&(0.49)&439&0.86\\
GSC 06785-00476&G7&1.56&(M8)&(0.3)&914&0.19\\
GSC 06204-01067&M2&0.49&(M5.5)&(0.11)&361&0.23\\
GSC 06780-01061&M3&0.36&(M5)&(0.12)&218&0.33\\
GSC 06784-00039&G7&1.56&(M7.5)&(0.05)&1962&0.03\\
GSC 06784-00997&M1&0.6&(M6)&(0.07)&697&0.11\\
GSC 06213-00306&K5&0.87&(M4.5)&(0.17)&461&0.2\\
GSC 06793-00868&M1&0.6&(M3)&(0.39)&291&0.65\\
GSC 06793-00806&M1&0.6&(M3.5)&(0.31)&274&0.51\\
RXJ1555.8-2512&G3&1.65&(M0)&(0.73)&1292&0.44\\
RXJ1555.8-2512&G3&1.65&(M5.5)&(0.1)&2118&0.06\\
RXJ1558.1-2405&K4&0.95&M5(M4)&0.13(0.22)&2632&0.23\\
RXJ1558.8-2512&M1&0.6&(M5)&(0.14)&1646&0.23\\
RXJ1559.2-2606&K2&1.12&(M2)&(0.48)&429&0.43\\
RXJ1600.7-2343&M2&0.49&(M2)&(0.48)&204&0.97\\
RXJ1602.8-2401B&K4&0.95&(M5)&(0.13)&1047&0.14\\
RXJ1604.3-2130&K2&1.12&M2(K5)&0.49(0.9)&2352&0.8\\
RXJ1606.6-2108&M1&0.6&(M1.5)&(0.54)&170&0.9\\
SCH16151115-24201556&M6&0.074&(M4.5)&(0.17)&2604&2.3\\
ScoPMS008a&M4&0.24&(M9)&(0.01)&229&0.05\\
ScoPMS016&M0.5&0.64&(M1.5)&(0.52)&199&0.82\\
ScoPMS042b&M3&0.36&(M7)&(0.06)&664&0.17\\
ScoPMS048&K0&1.35&(K1)&(1.27)&442&0.94\\
ScoPMS052&K0&1.35&(M2.5)&(0.45)&2764&0.33\\
USco80&M4&0.24&(K5)&(0.89)&1779&3.7\\
USco-155532.4-230817&M1&0.6&(M3.5)&(0.28)&260&0.46\\
USco-160202.9-223613&M0&0.68&(M4)&(0.25)&334&0.37\\
USco-160258.5-225649&M2&0.49&(M7)&(0.06)&175&0.13\\
USco-160428.4-190441&M3&0.36&M4(M5.5)&0.24(0.1)&1417&0.27\\
USco-160611.9-193532&M5&0.13&(M6)&(0.07)&1563&0.5\\
USco-160700.1-203309&M2&0.49&(M6)&(0.78)&1689&1.6\\
USco-160702.1-201938&M5&0.13&(M5.5)&(0.1)&236&0.77\\
USco-160822.4-193004&M1&0.6&K9(M3)&0.71(0.38)&1953&0.63\\
USco-160900.7-190852&K9&0.71&M5(M4.5)&0.13(0.18)&2743&0.25\\
USco-160904.0-193359&M4&0.24&(M5)&(0.15)&186&0.63\\
USco-160908.4-200928&M4&0.24&(M4.5)&(0.19)&280&0.8\\
USco-160936.5-184800&M3&0.36&(M6)&(0.07)&2896&0.19\\
USco-161031.9-191305&K7&0.77&(M7.5)&(0.04)&828&0.05\\
USco-161039.5-191652&M2&0.49&(M5.5)&(0.11)&2168&0.22\\
GSC06770-00655&K5&0.87&(M2.5)&(0.43)&1306&0.49\\
GSC06770-00655&K5&0.87&(M0)&(0.68)&4295&0.78\\
RXJ1524.2-3030A&K0&1.35&M1(K7)&0.60(0.76)&2926&0.56\\
RXJ1529.4-2850&G8&1.52&(G9)&(1.44)&357&0.95\\
RXJ1529.4-2850&G8&1.52&(G8)&(1.47)&300&0.97\\
RXJ1530.4-3218&G7&1.56&(G8)&(1.47)&300&0.94\\
RXJ1536.5-3246&M3&0.36&(M3.5)&(0.3)&344&0.82\\
RXJ1537.0-3136&G7&1.56&K7(G8)&0.77(1.48)&779&0.95\\
RXJ1539.4-3446B&K7&0.77&(M7)&(0.06)&4031&0.08\\
RXJ1539.4-3446B&K7&0.77&($>$M9)&($<$0.01)&4259&$<$0.02\\
RXJ1539.4-3446B(/C)&K7&0.77&M2(M4)&0.49(0.22)&922&0.29\\
RXJ1540.7-3121&M4&0.24&M5(M4.5)&0.13(0.21)&863&0.89\\
RXJ1543.8-3306&M3&0.36&(M3.5)&(0.29)&405&0.8\\
RXJ1545.2-3417&K0&1.35&(M0)&(0.7)&377&0.52\\
RXJ1554.0-2920&M0&0.68&(M4)&(0.23)&209&0.34\\
RXJ1554.0-2920&M0&0.68&(M4.5)&(0.18)&3818&0.26\\
\enddata
\tablenotetext{a}{Values in parentheses are estimated from the system flux
ratio $\Delta$$J$ and the spectroscopically determined properties of the         
primary.}
\tablenotetext{b}{Estimated statistical uncertainties are $\sim$10\% for
mass ratios, $\sim$20\% for secondary masses, $\sim$2-3 subclasses for
spectral types, and $\sim$10\% for projected separations.}
\end{deluxetable*}

\end{document}